\documentclass[twocolumn,english,prx,reprint,superscriptaddress]{revtex4-1}
\usepackage[T2A,T1]{fontenc}
\usepackage[utf8]{inputenc}
\usepackage{verbatim}
\usepackage{amsmath}
\usepackage{amssymb}
\usepackage{graphicx}
\usepackage{color}
\usepackage{braket}

\usepackage{soul}
\setstcolor{red}

\makeatletter

\usepackage{graphicx} 
\graphicspath{{plots_submit/}} 

\usepackage[colorlinks,citecolor=blue,linkcolor=red,urlcolor=blue]{hyperref}

\usepackage{amsthm}
\usepackage{braket}

\makeatother

\usepackage{babel}
\begin{document}
\title{Probing  infinite many-body quantum systems with finite-size quantum simulators}

\author{Viacheslav Kuzmin}
\affiliation{Center for Quantum Physics, University of Innsbruck, Innsbruck A-6020, Austria}
\affiliation{Institute for Quantum Optics and Quantum Information of the Austrian Academy of Sciences, Innsbruck A-6020, Austria}

\author{Torsten V. Zache}
\affiliation{Center for Quantum Physics, University of Innsbruck, Innsbruck A-6020, Austria}
\affiliation{Institute for Quantum Optics and Quantum Information of the Austrian Academy of Sciences, Innsbruck A-6020, Austria}

\author{Christian Kokail}
\affiliation{Center for Quantum Physics, University of Innsbruck, Innsbruck A-6020, Austria}
\affiliation{Institute for Quantum Optics and Quantum Information of the Austrian Academy of Sciences, Innsbruck A-6020, Austria}

\author{Lorenzo Pastori}
\affiliation{Center for Quantum Physics, University of Innsbruck, Innsbruck A-6020, Austria}
\affiliation{Institute for Quantum Optics and Quantum Information of the Austrian Academy of Sciences, Innsbruck A-6020, Austria}

\author{Alessio Celi}
\affiliation{Center for Quantum Physics, University of Innsbruck, Innsbruck A-6020, Austria}
\affiliation{Institute for Quantum Optics and Quantum Information of the Austrian Academy of Sciences, Innsbruck A-6020, Austria}
\affiliation{Departament de Física, Universitat Autònoma de Barcelona, 08193 Bellaterra, Spain}

\author{Mikhail Baranov}
\affiliation{Center for Quantum Physics, University of Innsbruck, Innsbruck A-6020, Austria}
\affiliation{Institute for Quantum Optics and Quantum Information of the Austrian Academy of Sciences, Innsbruck A-6020, Austria}    

\author{Peter Zoller}
\affiliation{Center for Quantum Physics, University of Innsbruck, Innsbruck A-6020, Austria}
\affiliation{Institute for Quantum Optics and Quantum Information of the Austrian Academy of Sciences, Innsbruck A-6020, Austria}

\begin{abstract}
    Experimental studies of synthetic quantum matter are necessarily restricted to approximate ground states prepared on finite-size quantum simulators. In general, this limits their reliability for strongly correlated systems, for instance, in the vicinity of a quantum phase transition (QPT). Here, we propose a protocol that makes optimal use of a given finite-size simulator by directly preparing, on its bulk region, a mixed state representing the reduced density operator of the translation-invariant infinite-sized system  of interest. This protocol is based on coherent evolution with a local deformation of the system Hamiltonian. For systems of free fermions in one and two spatial dimensions, we illustrate and explain the underlying physics, which consists of quasi-particle transport towards the system's boundaries while retaining the bulk ``vacuum''. For the example of a non-integrable extended Su-Schrieffer-Heeger model, we demonstrate that our protocol enables a more accurate study of QPTs. In addition, we demonstrate the protocol for an interacting spinful Fermi-Hubbard model with doping for 1D chains and a small two-leg ladder, where the initial state is a random superposition of energetically low-lying states.
\end{abstract}

\maketitle

\section{Introduction}

\begin{figure}[!ht]
    \centering
    \includegraphics[scale=1]{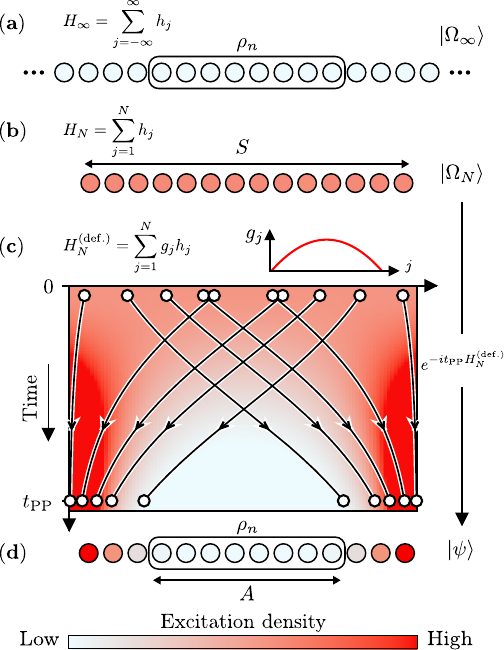}
    \caption{
    \label{fig:overview}
    Overview of Purification Preparation (PP), illustrated for a one-dimensional chain (see text). 
    \textbf{(a)} Ground state $\ket{\Omega_\infty}$ on an infinite-sized lattice with translation-invariant Hamiltonian $H_\infty$ composed of local operators $h_j$.
    \textbf{(b)} The ground state $\ket{\Omega_N}$ for a finite-size system $S$ can be interpreted as a part of $\ket{\Omega_\infty}$, populated with low-energy excitations (quasi-particles).
    \textbf{(c)} Density and trajectories of excitations, represented by circles, during coherent evolution with deformed Hamiltonian $H^{(\text{def.})}_N$, parametrized by $g_j$, on the quantum simulator (obtained in the continuum limit of free-fermion system for evolution with a {\em parabolically deformed} Hamiltonian, see Appendix~\ref{app:continuum}). At time $t_\text{PP}$ the initial state $\ket{\Omega_N}$ is transformed into a purification $|\psi\rangle$.
    \textbf{(d)} The final state $|\psi\rangle$ with subsystem $A$ that corresponds to the part of $\ket{\Omega_\infty}$.
    White to red colors indicate the density of excitations (colorbar at the bottom).
    } 
\end{figure}

A major goal of quantum simulation is the study of ground states (GSs) of quantum many-body systems with  phase transitions emerging in the thermodynamic limit.
Aiming for accurate characterization of GS phase diagrams raises the question of how to directly probe -- {\it on a finite-size quantum simulator} -- the GS $|\Omega_\infty\rangle$ of an {\em infinite} system described by a translation-invariant Hamiltonian \mbox{$H_\infty 
$}, especially in the vicinity of and at critical points [see Fig.~\ref{fig:overview}(a)].
In practice, any quantum simulator will necessarily be limited to a finite size, say $N^d$ sites for a $d$-dimensional lattice system. In this context, we note the remarkable recent  experimental progress~\citep{NAP25613,altman2021quantum} culminating in the realization of almost perfectly isolated synthetic quantum systems consisting of tens or hundreds of atoms as Hubbard or spin lattice models in various spatial dimensions~\cite{Browaeys2020, Ebadi2021,Scholl2021,Mazurenko2017,Kokail2019,Koepsell2019,Gross2017,Sun2021,Vijayan2020,Holten2021,Nichols2019,Brown2019,Monroe2021,Semeghini2021}. 
The restriction to finite size is less relevant as long as the intrinsic correlation length $\xi$ of the target state $|\Omega_\infty\rangle$ is sufficiently small compared to the available system size, i.e., $\xi \ll N$ (with $\xi$ in units of the lattice spacing). 
In this case, an approximation of the finite-size GS $|\Omega_N\rangle$ of \mbox{$H_N
$} [system $S$, see Fig.~\ref{fig:overview}(b)], which might have been prepared via adiabatic state preparation or variational techniques~\cite{Kokail2019,Ebadi2021}, allows for a faithful study of $|\Omega_\infty\rangle$. 
However, if the correlation length diverges, $\xi\rightarrow \infty$, such as at a quantum phase transition (QPT), significant finite-size effects destroy a faithful representation of the infinite-size GS on a finite system with $N \leq \xi$.

In this paper we resolve this situation by considering {\em pure} many-body states $|\psi\rangle $, living on a finite-size quantum simulator $S$ of linear dimension $N$, such that its reduced density operator $\rho_n$ for a subsystem $A$ of size $n<N$ equals the {\em mixed} state represented by the reduced state of the desired infinite GS $|\Omega_\infty\rangle$ [see Fig.~\ref{fig:overview}(d)],
\begin{align}\label{eq:purification}
\rho_n =\text{Tr}^{(S)}_{\neg A}\left[|\psi\rangle\langle\psi|\right]=\text{Tr}^{(\infty)}_{\neg A }\left[\ket{\Omega_\infty}\bra{\Omega_\infty}\right].
\end{align}
Here the traces $\text{Tr}^{(S)}_{\neg A}$ and $\text{Tr}^{(\infty)}_{\neg A}$ are taken over the complement of $A$ in the finite and infinite systems, respectively. On a formal level, we note that quantum information theoretical considerations guarantee the existence of a purification $|\psi\rangle $ of $\rho_n$  for the case $N=2n$~\cite{nielsen2002quantum}, with $n$ lattice sites representing the system of interest $A$, and the remaining $n$ sites playing the role of an auxiliary system, or `reservoir'.  The challenge to be addressed in a quantum simulation setting is to find specific, experimentally realizable protocols which allow the preparation of the purifications $|\psi\rangle $ for properly chosen $n$ for given $N$. Below we will refer to such protocols as Purification Preparation (PP). 

The central result of this paper is that Purification Preparation can be achieved in analog quantum simulation setups as engineered quench dynamics with {\em spatially deformed} system Hamiltonians, something which is readily implemented in today's experiments. We consider a situation where the analog quantum simulator  provides us not only with an implementation of the (in bulk translationally invariant) Hamiltonian of the form \mbox{$H_N = \sum_{j \in S} h_j$}, but also allows the realization of {\em  spatially deformed} Hamiltonians \mbox{$H_N^{(\text{def.})} = \sum_{j\in S}g_{j} h_{j}$} with $\{g_{j} \}$ a given spatial pattern of local couplings \cite{kokail2021quantum}. In its simplest form, the protocol starts with the preparation of a pure state approximating the GS of the finite system,  $\ket{\psi(0)} \approx \ket{\Omega_N}$, with \mbox{$H_N \ket{\Omega_N} = E \ket{\Omega_N}$}. Our claim is that the coherently evolved state \mbox{$\ket{\psi(t)}= e^{-iH_N^{(\text{def.})}t}\ket{\psi(0)}$} provides an approximate purification $\ket{\psi}$ for a proper choice of deformation parameters $\{g_j\}$ (in particular in form of a parabolic deformation) and evolution time $t = t_\text{PP}$, i.e.~$\ket{\psi}=\ket{\psi(t_\text{PP})}$. 

While the present paper will provide detailed reasoning behind the functioning of the protocol from various perspectives, which includes connections with recent studies of quench dynamics within conformal field theory (CFT)~\cite{Wen2018b,Wen2018,Fan2020_SSD,PhysRevResearch.3.023044,Fan2021} and numerous numerical illustrations, we outline in Fig.~1 a simple physical picture behind PP as engineered quasi-particle dynamics. Here we interpret the initial state $\ket{\psi(0)} \approx \ket{\Omega_N}$ in terms of excitations above the target state, and the physical essence of the dynamical evolution with the parabolically deformed Hamiltonian consists of ``cleaning'' the bulk $A$ from these excitations by moving them to, and accumulating them at the edge $S\diagdown A$ [Fig.~\ref{fig:overview}(c)], and thus effectively ``cooling'' the bulk.
In this way the PP protocol essentially eliminates boundary effects and restores translation invariance in the bulk $A$. 
We demonstrate our protocol for non-interacting fermions in one and two spatial dimensions as well as an interacting spin chain, a 1D Fermi-Hubbard model and preliminary results for a Fermi-Hubbard model on a two-leg ladder. These examples suggest broad applicability of our approach to both interacting and non-interacting systems in different spatial dimensions.
While the present work  focuses on initial {\em pure} states in PP, we emphasize that the approach is readily  applied to mixed {\em thermal} states, where again the bulk is ``cooled''  by accumulating  thermal excitations dominantly at the edge.

Finally, in connecting the present theoretical work to a feasible experimental protocol,  {\em tools for verification}, quantifying the approach to the target state $\rho_n$, must be developed. This can be achieved via recently developed techniques for Entanglement Hamiltonian (EH) learning ~\cite{carrasco2021theoretical, kokail2021entanglement,kokail2021quantum, qi2019determining,Bairey2019,zhu2019reconstructing}. Writing $\rho_n \sim \exp (-\tilde H_n)$ with $\tilde H_n$ defining EH, the Bisognano-Wichmann theorem of CFT~\citep{Bisognano_1975, Bisognano_1976,Hislop1982,Casini2011} applied to lattice models (see, e.g.,  \cite{Dalmonte2018,kokail2021entanglement,kokail2021quantum}) predicts that the EH for ground states has the simple structure of a parabolically deformed system Hamiltonian. EH learning thus provides the techniques to monitor and thus verify the approach to the target state $\rho_n$ in PP. Again we will illustrate this with examples.

The structure of this paper is as follows.
We first provide an overview of our results in Sec.~\ref{sec:teaser} where we introduce our state preparation protocol and present the results of its application to an interacting spin chain, demonstrating the potential of our approach for mapping out the GS phase diagram. In the following Sec.~\ref{sec:Protocol}, we discuss the physics 
underlying our protocol in more detail. For the examples of non-interacting fermions in one and two spatial dimensions, we provide extensive numerical evidence supplemented by analytical predictions to explain how the desired purification is realized. Subsequently, we return to the interacting spin chain in Sec.~\ref{sec:int_spinchain} and demonstrate that our protocol enables a more accurate determination of universal critical exponents that characterize a QPT and the underlying CFT. In Sec.~\ref{sec:hubbard} we demonstrate the protocol for an interacting spinful Fermi-Hubbard model with doping for 1D chains, and further results for two-leg ladders are presented in Appendix~\ref{app:Hubbard}. In this study the initial state are taken as random superposition of energetically low-lying states, i.e. effectively mixed initial states. We conclude our work with a brief discussion in Sec.~\ref{sec:discussion}. The appendices contain additional details about our protocol, its analysis, and our numerical and analytical calculations.

\section{\label{sec:teaser}Overview of concepts and results}

This section will provide an overview of the concepts and application of the Purification Preparation protocol. We will first describe and motivate the  protocol {\em per se}, followed by the illustrative example of the study of the phase diagram of the extended Su-Schrieffer-Heeger model, comparing the case of finite vs.~infinite system.  

\subsection{Purification Preparation Protocol}

We consider  a quantum simulator with finite linear size $N$, and we assume here
pure initial states $\ket{\psi(0)}$, which are close to the GS $\ket{\Omega_N}$ of a translation-invariant finite-size local Hamiltonian \mbox{$H_N$}. In a typical quantum simulation experiment $\ket{\Omega_N}$ corresponds to a GS with open boundary conditions (OBC). For simplicity of writing, we describe here the case of a finite one-dimensional (1D) chain with OBC (see Fig.~\ref{fig:overview}), where \mbox{$H_N = \sum_{j=1}^{N-1} h_{j,j+1}$} and $h_{j,j+1}$ couple neighboring lattice sites in the total system $S$ with $N$ sites. Our goal is to realize the reduced state $\rho_n$ [Eq.~\eqref{eq:purification}] on $n<N$ sites in the middle region $A$.

Our protocol to prepare the purification $\ket{\psi}$ in Eq.~\eqref{eq:purification} as $\ket{\psi(t_\text{PP})}$ consists of two main steps: 
\begin{enumerate}
    \item Evolution of the initial state $\ket{\psi(0)}$ with a deformed Hamiltonian $H_N^{(\text{def.})}$,
    \mbox{$\ket{\psi(t)} = e^{-iH_N^{(\text{def.})}t}\ket{\psi(0)}$},
    for some time $t$. As we argue below, a good choice for models with dynamical critical exponent $z=1$ is 
    \begin{align}
    H_N^{(\text{def.})} = \sum_{j=1}^{N-1}g_j h_{j,j+1}\;, && g_j = \frac{(N-j)j}{(N/2)^2}\label{parabolic_def}.
    \end{align}
    corresponding to a {\em parabolic} deformation of the system Hamiltonian 
    \item Finding the optimal time $t_\text{PP}$ to stop the evolution as the time that corresponds to the first maximum of the second R\'{e}nyi entropy (purity) of $\ket{\psi(t)}$ measured for the edge region $E \subset S \diagdown A$ of the simulator $S$.
\end{enumerate}

Let us briefly comment on the rationale behind these steps (for more details, see Sec.~\ref{sec:Protocol}).
The main idea relies on interpreting the initial state $\ket{\psi(0)}$ as a subsystem of the infinite system in some excited state. 
More formally, this can be formulated as follows: According to  Kraus' theorem~\citep{Kraus1983}, the initial state can be written as 
\[
\ket{\psi(0)} \bra{\psi(0)}=\sum_{\alpha}\mathrm{Tr}^{(\infty)}_{\neg S}[ \hat{O}_{\alpha}\ket{\Omega_\infty}\bra{\Omega_\infty} \hat{O}_{\alpha}^{\dagger}],
\]
where the operators $\hat{O}_\alpha$ act non-trivially only on the finite system $S$, viewed as a subsystem on an infinite system, creating there excitations above the ground state $\ket{\Omega_\infty}$.

The first step of our protocol implements an evolution to 
\[\ket{\psi(t)} \bra{\psi(t)}= \sum_{\alpha}\mathrm{Tr}^{(\infty)}_{\neg S}[ \hat{O}_{\alpha}(t)\ket{\Omega_\infty (t)}\bra{\Omega_\infty(t)} \hat{O}_{\alpha}^{\dagger}(t)]\]
in such a way that the support of 
the evolved operators 
\[\hat{O}_{\alpha}(t) = e^{-iH_N^{(\text{def.})}t}\, \hat{O}_{\alpha}\, e^{iH_N^{(\text{def.})}t}\] in the subsystem $A \subset S$ decreases with time $t$~\footnote{To determine the evolution with $H_N^{(\text{def.})}$ for objects in the infinite system, we extend it as acting trivially outside $S$.}.
This describes transport of the excitations from the bulk towards the edge of the simulator (see Sec.~\ref{sec:mode_mapping} for an explicit demonstration of this transport).
Ideally, we want that at time $t_\text{PP}$ $\hat{O}_{\alpha}(t)$ has no more effect on $A$, such that 
\[\text{Tr}^{(S)}_{\neg A}[\ket{\psi(t_\text{PP})} \bra{\psi(t_\text{PP})}] = \text{Tr}^{(\infty)}_{\neg A}[\ket{\Omega_\infty (t_\text{PP})}\bra{\Omega_\infty(t_\text{PP})}].\]
If additionally $\text{Tr}^{(\infty)}_{\neg S}[\ket{\Omega_\infty }\bra{\Omega_\infty}]$ commutes with $H_N^{(\text{def.})}$, i.e., if the ``bulk vacuum'' is preserved by the evolution, we obtain the desired result \mbox{$\text{Tr}^{(S)}_{\neg A}[\ket{\psi(t_\text{PP})} \bra{\psi(t_\text{PP})}] =\rho_n$}. This preservation is guaranteed if the deformed Hamiltonian is proportional to the Entanglement Hamiltonian (EH)
\[\tilde{H}_N = - \log \left\{ \text{Tr}^{(\infty)}_{\neg S}\left[\ket{\Omega_\infty }\bra{\Omega_\infty} \right]\right\} \;.\] In this context, the Bisognano-Wichmann theorem~\citep{Bisognano_1975, Bisognano_1976} for a CFT~\cite{Hislop1982,Casini2011}, which
states that the EH is given by a parabolic deformation, suggests our choice of Eq.~\eqref{parabolic_def}.

The criterion used in the second step to determine the optimal time $t_\text{PP}$ is motivated by the spreading of entanglement due to the propagation of entangled quasi-particle pairs~\cite{calabrese2005evolution}. We expect the (von Neumann) entanglement entropy $S_{S \diagdown A}$ of the boundary region $S \diagdown A$, which equals the entropy $S_A$ of the bulk for an initial pure state,
\begin{align}
    S_{S \diagdown A} =  S_A = - \text{Tr} \left[\rho_n \log \rho_n\right] \;,
\end{align}
to grow while excitations are transported out of $A$. Eventually, these excitations will reflect at the boundary of $S$ and return into the bulk $A$. We use this to estimate the optimal evolution time $t_\text{PP}$ as the first maximum of the entropy of an edge region $E\subset S\diagdown A$, which for a spatially symmetric system we take as a few sites on one side of the chain, described by state $\rho_E$. Further, we choose to measure the second R\'{e}nyi entropy 
\begin{align}
    S^{(2)}_E = - \log \left\{\text{Tr}\left[\rho_E^2\right]\right\} \;,
\end{align}
as an experimentally accessible~\citep{Brydges2019} alternative for 
the von Neumann entropy of the edge region $E$. A more accurate way to monitor the build-up of the entanglement structure corresponding to the target reduced state $\rho_n$ is discussed further below in section~\ref{section:EHapproach_for_tPP}.

Before showing the first example of our protocol, we would like to briefly discuss its experimental requirements. Both state preparation, e.g., adiabatically or variationally, as well as quench dynamics, are nowadays routinely implemented in quantum simulation experiments. The main challenge of our protocol thus consists of realizing a deformation of the system Hamiltonian. This requires a certain degree of programmability, i.e., we explicitly assume the capability of tuning local Hamiltonian parameters. Alternatively, the corresponding time evolution can also be realized stroboscopically. For completeness, we show an explicit example of such a ``Trotterized'' approach in Appendix~\ref{app:Trotter}.

\subsection{Application: phase diagram of extended Su-Schrieffer-Heeger model}

\begin{figure*}[!ht]
\centering \includegraphics[]{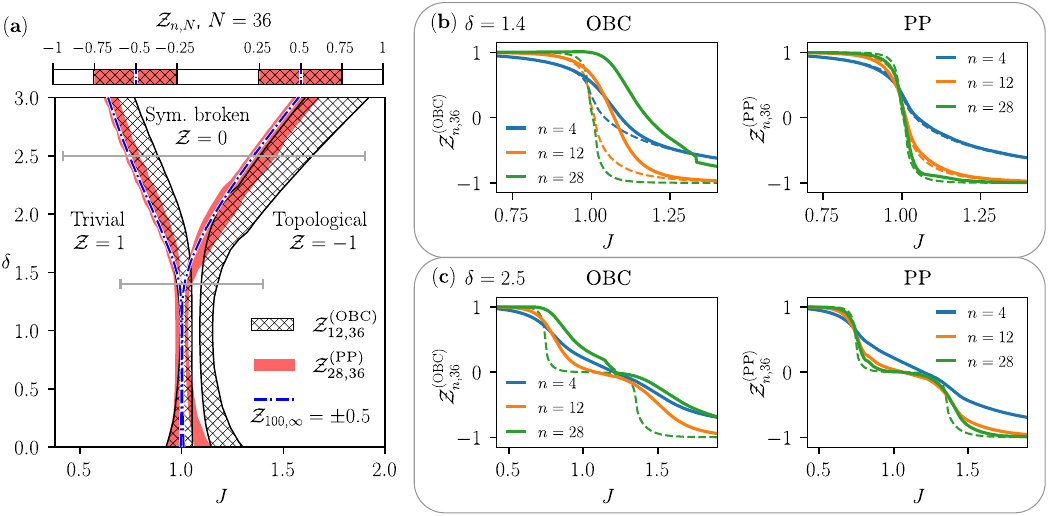} \caption{
Phase transitions
in the extended SSH model~[Eq.~\eqref{eq:SSH}]
with $N=36$ sites. 
\textbf{(a)} Phase diagram obtained using the partial reflection invariant $\mathcal{Z}_{n,N}$, [Eq.~\eqref{eq:MBTI}]. We use the values  $-0.5$ and $0.5$ to identify the phase transitions in the infinite system (dashed-dotted lines obtained for a subsystem of $n=100$ sites) and the uncertainty intervals $[-0.75,-0.25]$ and $[0.25,0.75]$ for the finite systems. For the latter, the corresponding uncertainty regions on the phase diagram are shown as the hatched and continuously shaded areas for $\mathcal{Z}_{n,N}^{(\text{OBC})}$ and $\mathcal{Z}_{n,N}^{(\text{PP})}$, respectively. 
The gray horizontal lines at $\delta=1.4$
and $\delta=2.5$ indicate the ranges of $J$ detailed in (b) and (c). 
\textbf{(b)} Dependencies of the $\mathcal{Z}_{n,N}$ for the OBC GSs (left panel) and PP states (right panel) on $J$ for $\delta=1.4$ and several values of $n$.
The dashed lines show the target values for the subsystems of the infinite system with the same values of $n$.
\textbf{(c)} The same as in (b) but for $\delta=2.5$.
}
\label{fig:string_order}
\end{figure*}

Before entering a detailed technical discussion of our protocol, we find it worthwhile to present results illustrating the power and potential of our approach. We focus on obtaining a quantum phase diagram obtained from observables measured in $A$ for the purification $\ket{\psi(t_\text{PP})}$. Specifically, we study an extended Su-Schrieffer-Heeger~\cite{SSH,Elben2020} (SSH) model on a system $S$ with an even number of sites $N$ and OBC, described by the 1D Hamiltonian 
\begin{multline}
H_N=\sum_{j=1}^{N/2}\left(\sigma^x_{2j-1}\sigma^x_{2j}+\sigma^y_{2j-1}\sigma^y_{2j}+\delta \sigma^z_{2j-1}\sigma^z_{2j}\right)\\
+J\sum_{j=1}^{N/2-1}\left(\sigma^x_{2j}\sigma^x_{2j+1}+\sigma^y_{2j}\sigma^y_{2j+1}+\delta \sigma^z_{2j}\sigma^z_{2j+1}\right),\label{eq:SSH}
\end{multline}
where $\{\sigma^x_{j},\sigma^y_{j},\sigma^z_{j}\}$ denote the Pauli operators at sites
$j$.

The phase diagram of this model, which is for $\delta\ne\{0,1\}$ is generally non-integrable, is shown in Fig.~\ref{fig:string_order}(a).
In the GS $\ket{\Omega_N}$, the symmetry-protected topological (SPT) phase is detected by measuring highly non-local many-body correlations, such as the partial reflection invariant~\citep{Elben2020}
\begin{equation}
\mathcal{Z}_{n,N}=\frac{\text{Tr}[\rho_{A}\mathcal{R}_{A}]}{\sqrt{\left(\text{Tr}[\rho_{A_{L}}^{2}]+\text{Tr}[\rho_{A_{R}}^{2}]\right)/2}},\label{eq:MBTI}
\end{equation}
in the bulk subsystem $A$ of $n<N$ sites ($n$ is even) centered in the middle of the system. 
Here, $\rho_A$ (and analogously $\rho_{A_{L/R}}$) is the reduced density matrix of subsystem $A=A_{L}\cup A_{R}$ with $A_{L}$ and $A_{R}$ the corresponding left and right partitions of $A$ with $n/2$ sites. The operator
$\mathcal{R}_{A}$
interchanges $A_{L}$ and $A_{R}$ with
respect to the reflection center.
In the thermodynamic limit, sending first $N\to\infty$ and then $n\to\infty$ with
$n$ being a multiple of four, $\mathcal{Z}_{n,N}$ approaches
quantized values: $\mathcal{Z}_{n,N}\to1$ for the trivial phase, $\mathcal{Z}_{n,N}\to-1$ for the SPT, and $\mathcal{Z}_{n,N}\to0$ for the symmetry-broken antiferromagnetic (AF) phase.

Since the typical subsystem size $n$ required to achieve
convergence for $\mathcal{Z}_{n,N}$ is determined by the correlation length $\xi$
~\citep{Elben2020}, increasingly large $n$ becomes necessary when approaching a phase transition. At the same time, in order to avoid boundary effects in the OBC GS $\ket{\Omega_N}$, $n$ must be taken relatively small compared to the system size $N$. We, therefore, expect a significant advantage in determining the phase boundaries on a finite system when using, instead of $\ket{\Omega_N}$, the state $\ket{\psi(t_\text{PP})}$ prepared by our protocol.

This expectation is confirmed in Fig.~\ref{fig:string_order}, where we present the results for the phase diagram obtained using a numerical simulation of our protocol:
the order parameter $\mathcal{Z}_{n,N}^{(\text{PP})}$ for the state $\ket{\psi(t_\text{PP})}$ with \mbox{$\ket{\psi(0)} = \ket{\Omega_N}$} is shown in comparison to $\mathcal{Z}_{n,N}^{(\text{OBC})}$ for the OBC GS $\ket{\Omega_N}$, and $\mathcal{Z}_{n,\infty}$ for the infinite system. The finite systems are simulated for $N=36$ using MPS techniques \citep{Schollwock2011}: DMRG
for finding $\ket{\Omega_N}$ and Trotter gates
for the time evolution resulting in $\ket{\psi(t_\text{PP})}$. The infinite case is treated with iDMRG~\citep{Schollwock2011, Orus2014} for details).
The dependencies of $\mathcal{Z}_{n,N}^{(\text{PP})}$ and $\mathcal{Z}_{n,N}^{(\text{OBC})}$ on $n$ for $\delta=1.4$ and
$\delta=2.5$ are shown in Figs.~\ref{fig:string_order}(b and (c). We observe that the convergence for the OBC GS [(b)] stops for $n\gtrsim N/3$ due to boundary effects, resulting in a large uncertainty in identifying the transition points. 
In contrast, due to suppression of the boundary effects, our protocol allows for larger values of $n$ and provides the results [(c)] which are much closer to those of the infinite system.
This enables much more precise identification of the phase transitions and leads to a quantitatively improved phase diagram with sharp and accurate phase boundaries, which are in very good agreement with the thermodynamic limit results, as demonstrated in Fig.~\ref{fig:string_order}(a) for $n=28$. Further below in section~\ref{sec:int_spinchain}, we will return to the present model and show that also critical exponents, which characterize the phase transitions, can be extracted more accurately using the PP protocol.

\section{Detailed analysis for non-interacting fermions\label{sec:Protocol}}

\begin{figure*}
\centering \includegraphics[width=7in]{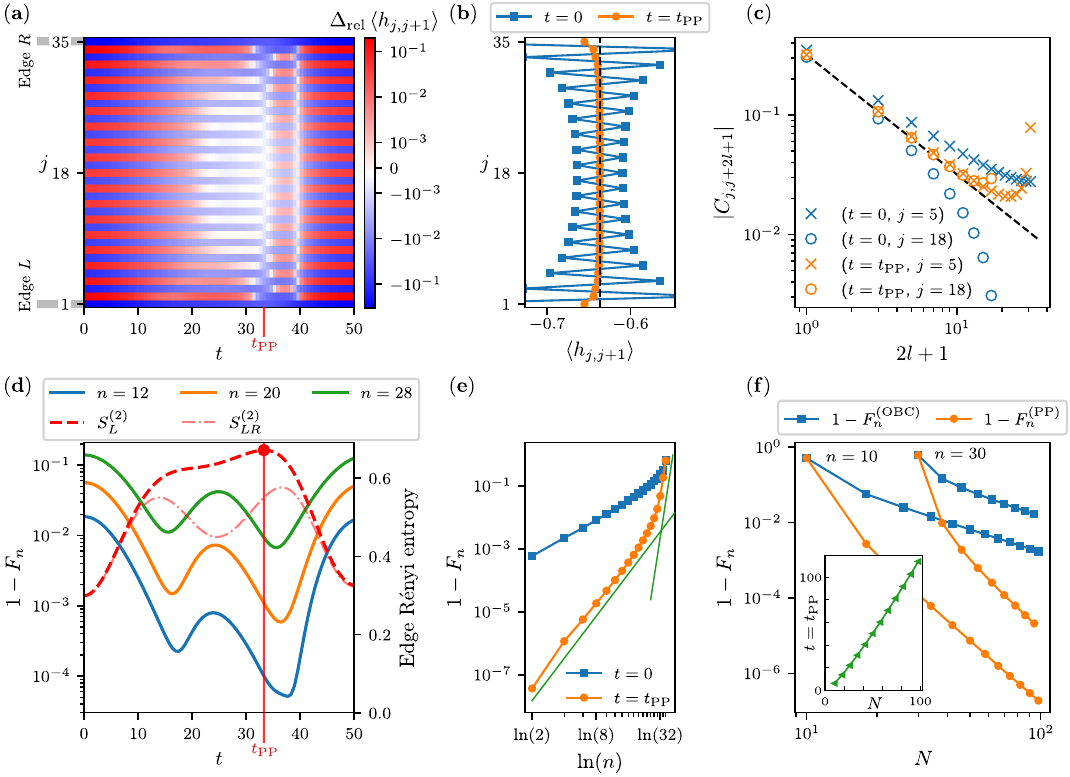} \caption{
Purification preparation for free fermions [see Eq.~(\ref{eq:1d_Fermi})] on a one-dimensional lattice with $N=36$ sites.
\textbf{(a)} Relative deviation of the energy density [Eq.~\eqref{eq:rel_en_div}] during the PP. The grey boxes indicate the edge subsystems used to measure the second R\'{e}nyi entropies $S^{(2)}_L$ and $S^{(2)}_{LR}$, plotted in (d), of the edge $L$ and both edges $L$ and $R$, respectively. Each edge consists of two neighbouring qubits.
\textbf{(b)} Energy density and \textbf{(c)} two-point correlation function [Eq.~\eqref{eq:cc_cors}] at times $t=0$ and $t=t_{\text{PP}}$. The black dashed lines correspond to the infinite GS.
\textbf{(d)} Infidelity [Eq.~\eqref{eq:inF}] for several bulk sizes $n$ (left axis)
and the second R\'{e}nyi entropies $S^{(2)}_L$ and $S^{(2)}_{LR}$ of the edges indicated in (a) (right axis). The red vertical line indicates the
time $t_{\text{PP}}$ at which $S^{(2)}_L$ reaches its maximum and the
PP protocol stops.
\textbf{(e)} Bulk infidelity [Eq.~\eqref{eq:inF}] versus the bulk size $n$ in the initial state ($t=0$) and at time $t_{\text{PP}}$. The green
lines indicate two distinct power-laws $\propto n^k$ (see text).
\textbf{(f)} Bulk infidelities $1-F_{n}^{(\text{OBC})}$
in the OBC GS and $1-F_{n}^{(\text{PP})}$ in the
PP state versus the system size $N$ for the two bulk sizes
$n=10$ and $n=30$. The inset shows the time $t_\text{PP}$ obtained from the maximum of the second R\'{e}nyi entropy $S^{(2)}_L$ of the two neighboring edge sites. 
}
\label{fig:PP_of_1D_fermi}
\end{figure*}

We now turn to an in-depth discussion of  our protocol. We start with a one-dimensional chain of non-interacting fermions at half filling described
by the Hamiltonian (throughout this article, we employ units where $H_N$ is dimensionless)
\begin{align}
H_{N}=\sum_{j=1}^{N-1} h_{j,j+1} \;, && h_{j,j+1} = \left(c_{j}^{\dagger}c_{j+1}+c_{j+1}^{\dagger}c_{j}\right)\;,\label{eq:1d_Fermi}
\end{align}
where $c^{(\dagger)}_{j}$ denote fermionic annihilation(creation) operators at site $j$. In the thermodynamic limit ($N\to\infty$), 
$H_{N}$ becomes a gapless model described by a low-energy effective theory that exhibits a relativistic dispersion with dynamical critical exponent $z=1$, namely the CFT of a free fermion. The goal of our protocol is to recover the properties of the GS $|\Omega_\infty\rangle$ of the infinite-size Hamiltonian, which has a translation-invariant energy density $\left\langle h_{j,j+1}\right\rangle $
 and two-point correlators 
 \begin{equation}
     |C_{j,j+2l+1}|\equiv|\langle c^\dagger_jc_{j+2l+1}\rangle|
     \label{eq:cc_cors}
 \end{equation}
 that decay algebraically with the spatial distance $|2l+1|$.

In the following, we consider a finite chain with $N=36$, for which Fig.~\ref{fig:PP_of_1D_fermi}(a) demonstrates the spatially modulated relative deviation of $\left\langle h_{j,j+1}\right\rangle$ from the exact value in the infinite GS, 
\begin{equation}
\Delta_{\text{rel}}\left\langle h_{j,j+1}\right\rangle \equiv\left(\left\langle h_{j,j+1}\right\rangle -\left\langle h_{j,j+1}\right\rangle _{\text{ex}}\right)/\left\langle h_{j,j+1}\right\rangle _{\text{ex}}
\label{eq:rel_en_div}
\end{equation}
during evolution with a parabolically deformed $H_N^{(\text{def.})}$. As shown in Fig.~\ref{fig:PP_of_1D_fermi}(b), the finite-size and boundary corrections of the initial OBC GS $|\Omega_N\rangle$ [$t=0$ in Fig.~\ref{fig:PP_of_1D_fermi}(a)] are strongly decreased at $t_\text{PP}$. Similarly, the initial correlators $|C_{j,j+2l+1}|$ decay too fast or too slow depending on $j$ being odd or even [see Fig.~\ref{fig:PP_of_1D_fermi}(c)], while translation invariance in the bulk and the target algebraic decay of correlators is restored at $t_\text{PP}$. Below we explain how the PP protocol achieves these results for subsystems $A\subset S$ with size $n<N$.

\subsection{Physical explanation of the protocol}
As outlined in the previous section, we regard the state $|\Omega_N\rangle$
as a part of a mixture of excited infinite-size states $\hat{O}_\alpha\ket{\Omega_\infty}$. In the considered system, these excited states differ from the ground state by a number of particle and hole excitations (quasi-particles) with momenta close to the Fermi momentum, which is $p_F=\pi/2$ (or $3\pi/2$) in our units.
This situation can be described by wave packets of quasi-particles with momentum $p$ created by the operators $\tilde{c}^\dagger_{p} \sim\sum_{j=-\infty}^{j=\infty}e^{ijp}c^\dagger_{j}$ for $p>p_F$ and by the Hermitian conjugate operator $\tilde{c}_{p}$ for $p<p_F$, which populate $\ket{\Omega_\infty}$. In the infinite system, the quasi-particles have a dispersion $\epsilon(p)=\left| \cos{p}\right|$. For a small deviation $\Delta_{\text{rel}}\left\langle h_{j,j+1}\right\rangle$, 
the quasi-particles in the OBC GS $|\Omega_N\rangle$ correspond to low-energy excitations, for which $p\approx p_F$, and the dispersion takes a linear relativistic form  $v_F \left| p-p_F\right|$ with a constant group velocity \mbox{$v(p)=\partial\varepsilon\left(p\right)/\partial p=\pm v_F$} with $v_F$ being the Fermi velocity. The idea of our protocol is to realize a unitary evolution $\mathcal{U}=e^{-it_\text{PP}H_N^{(\text{def.})}}$ which moves these quasi-particles to the boundaries of $S$, without creating new excitations in the middle of the system. In the absence of quasi-particles, the bulk $A$ approximates a subsystem of the true vacuum, and ideally 
\[
\text{Tr}^{(S)}_{\neg A}\left[\mathcal{U}|\Omega_N\rangle\langle\Omega_N|\mathcal{U}^{\dagger}\right]=\text{Tr}^{(\infty)}_{\neg A}\left[|\Omega_\infty\rangle\langle\Omega_\infty|\right].
\]

This evolution can be achieved with a local deformation of the finite-size Hamiltonian, \mbox{$H_N^{(\text{def.})}=\sum_{j=1}^{N-1}g_{j}h_{j,j+1}$} with an appropriately chosen deformation $g_j$. A sufficiently smooth deformation locally modifies the velocity, $v_j\propto g_j$, such that the quasi-particles quickly leave
the bulk and spend a long time at the edge of the finite system before going back into the bulk (see Appendix~\ref{app:continuum} for a detailed derivation). At the same time, $g_j$ should be chosen such that the properties of the bulk vacuum are retained.
Intuitively, any deformation $g_j$ that smoothly changes from unity at the center of the bulk to zero at the system's edge should provide such dynamics
(see also the studies~\citep{Wen2018,Fan2020_SSD,Fan2021} of Floquet CFTs with deformed Hamiltonians). Since the bulk vacuum is conserved for evolution with a Hamiltonian proportional to the EH (or commuting with it), we expect that a quasi-local approximation of the EH will fulfill the desired conditions and provide the desired purification.

Returning to the chain of free fermions~[Eq.~\eqref{eq:1d_Fermi}], a parabolic deformation, $g_j= (N-j)j/(N/2)^{2}$,
is a quasi-local approximation~\citep{Eisler2017} of the EH Hamiltonian that also exactly commutes~\citep{Peschel2004} with the reduced density matrix \mbox{$\rho_N = \text{Tr}^\infty_{\neg S} \left[|\Omega_\infty \rangle \langle \Omega_\infty |\right]$} of the infinite GS, \mbox{$[ H^{(\text{def.})}_N,\rho_N] = 0$}.
Moreover, under evolution with this parabolic deformation any low-energy quasi-particle in the bulk reaches the edge of the system exponentially fast with a time-scale $t\propto N$ [Eq.~\eqref{eq:parbolic_trajectory} in Appendix~\ref{app:continuum} gives an analytical expression for the quasi-particle's trajectories, with several trajectories demonstrated in Fig.~\ref{fig:overview}(c)]. We note that a deformation function $g_j$ that connects more
smoothly to zero at the edge of the system, specifically the so-called sine-square
deformation that has been studied for CFTs~\citep{Wen2018b, Fan2020_SSD}, leads to an algebraic rather than an exponential approach to the edge (see Appendix~\ref{app:continuum}). 

To test our predictions, we simulated the evolution of the initial state $|\Omega_N\rangle$ under the deformed Hamiltonian
for $N=36$, (see Appendix~\ref{methods:FF_simulation} for details on the numerics). The resulting motion of quasi-particles can be tracked indirectly by monitoring the system's energy density during the evolution. After a certain evolution
time $t_{\text{PP}}$, the initial oscillations of the energy density are strongly suppressed,
and the bulk energy density exhibits translational invariance with
a value close to the target one of the true vacuum, as shown in Fig.~\ref{fig:PP_of_1D_fermi}(a) and Fig.~\ref{fig:PP_of_1D_fermi}(b).
At the same time, the two-point correlators $|C_{j,j+2l+1}|$
in the bulk, plotted in Fig.~\ref{fig:PP_of_1D_fermi}(c), demonstrate
the target algebraic decay.

More complex correlations in the bulk state \mbox{$\rho_n(t_\text{PP}) = \text{Tr}^{(S)}_{\neg A} \left[\ket{\psi(t_\text{PP})} \bra{\psi(t_\text{PP})}\right]$} also approach their
target values, which follows from the decrease of the bulk infidelity
\begin{equation}
1-F_{n}=1-\left(\text{Tr}_{A}\sqrt{\sqrt{\rho_n(t_\text{PP})}\rho_{n}\sqrt{\rho_{n}(t_\text{PP})}}\right)^2\label{eq:inF}
\end{equation}
with respect to the target reduced state \mbox{$\rho_n = \text{Tr}^{(\infty)}_{\neg A} [\ket{\Omega_\infty} \bra{\Omega_\infty}]$} of the infinite GS by several orders of magnitude, as demonstrated in Fig.~\ref{fig:PP_of_1D_fermi}(d) and Fig.~\ref{fig:PP_of_1D_fermi}(e).
The infidelity of the subsystem in the PP state, Fig.~\ref{fig:PP_of_1D_fermi}(e),
grows approximately as $n^{k}$ with $k$ changing to a larger value at $n\approx28$
indicating a ``concentration'' of the quasi-particles within a relatively small edge region, $\approx 4$ sites at each side. The same conclusion
can be deduced from the faster growth of the deviation of the energy density at these
edge sites [Fig.~\ref{fig:PP_of_1D_fermi}(b)] from its target value. Since the
infidelity is related to the density of the quasi-particles~\citep{Dziarmaga2010}, these numerical observations strongly support our arguments about the evolution with the parabolically deformed Hamiltonian.

In an experiment, it is hardly possible to measure the infidelity with respect to the true (Fermi sea) vacuum. Instead, we estimate the 
time $t_{\text{PP}}$ to stop the evolution by measuring the second R\'{e}nyi entropy~\citep{Brydges2019}
on a small subsystem at the system boundary. This is illustrated
in Fig.~\ref{fig:PP_of_1D_fermi}(d), where we plotted the entropy $S^{(2)}_L$ of the two left-most sites of the chain. The maximum of the entropy closely
matches the minima of the infidelities, which we attribute to the boundary region $S \diagdown A$
``storing'' the necessary entanglement required for the global pure
state to produce the correct mixed state in the bulk. For any finite lattice  spacing, quasi-particles
eventually reflect at the system's boundaries and return
into the bulk, thereby decreasing both the edge entropy as well as the bulk fidelity for times $t>t_{\text{PP}}$. We also note that the edge entropy provides experimentally more favorable criterion for determining the optimal time $t_{\text{PP}}$ than measuring small deviations of the energy density in the bulk from its target homogeneous value, see Fig.~\ref{fig:PP_of_1D_fermi}(b). 
This is because the values of the entropy, $S^{(2)}_L\sim 1$, and its rapid decrease for $t>t_{\text{PP}}$ provide much better visibility in the presence of experimental shot noise. Moreover, the entropy $S^{(2)}_{LR}$ corresponding to two edge sites on each end of the chain [also shown in Fig.~\ref{fig:PP_of_1D_fermi}(d)] indicates a similar maximum as $S^{(2)}_{L}$ with deviations that result from correlation among the two boundaries of the chain. In view of experimental feasibility (in particular for higher dimensions), we focus on the simpler edge entropy $S^{(2)}_{L}$ throughout the rest of this paper.

Finally, Fig.~\ref{fig:PP_of_1D_fermi}(f) presents the numerically obtained values of the time
$t_{\text{PP}}$ and the corresponding infidelity $1-F_{n}$ versus the total size
$N$ for the fixed subsystem sizes $n=10$ and $n=30$ before and after PP protocol. As expected
from the quasi-particle picture, we obtain $t_{\text{PP}}\sim N$.
The infidelities in both cases decrease approximately as $N^{-k}$,
where we find $k$ to be independent on the bulk size $n$. For the OBC GS we retrieve
$k^{(\text{OBC})}\approx2.0$ while for the PP states $k^{(\text{PP})}\approx5.4$. 
Therefore, our protocol on a simulator of size $N$ results in the same infidelity in the bulk as a simulator of the size $\sim N^{2.7}$ when using the OBC GS. 
We can further estimate the time required to prepare the corresponding states in both strategies: Using a non-linear adiabatic
sweep~\citep{Blundell2016}, one can prepare the OBC GS of size $N$ in a time $\sim N$. With the time $t_{\text{PP}}\sim N$ required for the PP protocol, this yields the total preparation time $\sim2N$ for the PP state of size $N$. To prepare the OBC GS which results in the same infidelity, on the other hand,
one needs time $\sim N^{2.7}$. These estimates demonstrate the advantages of our protocol for critical system studies.

\subsection{\label{sec:mode_mapping}Interpretation of quasi-particle motion as mode mapping}

\begin{figure}
\includegraphics{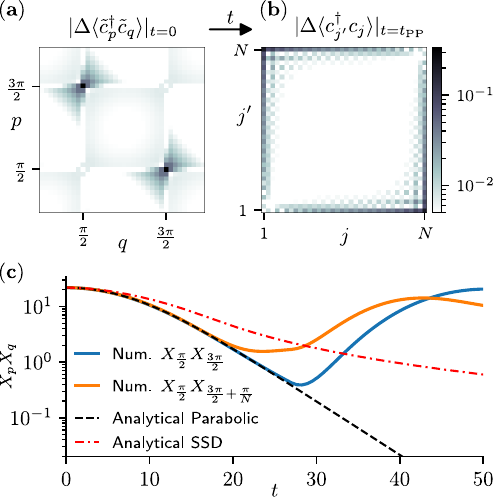}

\caption{Mode mapping from the momentum to the position space for the fermionic chain with size $N=36$ during the evolution under the parabolically deformed Hamiltonian. 
\textbf{(a)} Initial deviation of the correlator in the OBC GS from the target value in the momentum space, and 
\textbf{(b)} the deviation of the correlator in the position space at time $t_{\text{PP}}$ from Fig.~\ref{fig:PP_of_1D_fermi}(a).
\textbf{(c)} Contribution of quasi-particles with momenta $q,p$ to the two-point correlators in the
subsystem $A$ corresponding to the sites $\{5, \dots, 32\}$.
The black dashed and red dash-dotted lines show the analytical result [see the discussion around Eq.~\eqref{eq:continuum_parab2}] for the parabolic and SSD deformations, respectively.
}
\label{fig:correlators_mapping}
\end{figure}

\begin{figure}
\includegraphics{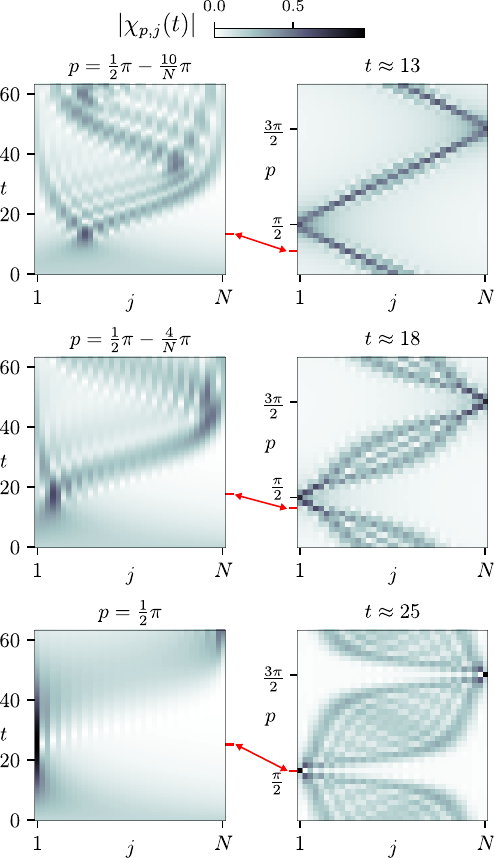}

\caption{Momentum-position mapping function $\left|\chi_{p,j}\left(t\right)\right|$,
Eq.~(\ref{eq:modes_mapping}), for fermionic chain with $N=36$.
Left column: Time evolution of $\left|\chi_{p,j}\left(t\right)\right|$ for several fixed values
of $p$. Right column: Values of $\left|\chi_{p,j}\left(t\right)\right|$
at fixed times corresponding to the ``collapse'' of the corresponding modes in the left column. Red arrows indicate identical data slices along spatial axis $j$ in the left and right columns.
}
\label{fig:modes_mapping}
\end{figure}

In the following, we give another view on
the physics that underlies the PP protocol for free fermions [Eq.~\eqref{eq:1d_Fermi}]. 
Specifically, we study how the PP dynamics maps finite-size fermionic modes in momentum space $\tilde{c}_{p}$,
\begin{align}
\tilde{c}_{p} & =\frac{1}{N}\sum_{j=1}^Ne^{-i\frac{2\pi j}{N}p}c_{j}.\label{eq:c_momentum}
\end{align}
to the modes in the position space $c_{j}$
In the Heisenberg picture, evolution with the deformed Hamiltonian
$H_N^{(\text{def.})}=\sum_{j'j}\tilde{h}_{j'j}c_{j'}^{\dagger}c_{j}$, with $\tilde{h}$
described by a Hermitian $N\times N$ matrix, acts on the fermion operators in position space as
\begin{align}
c_{j'}(t) & =e^{itH_N^{(\text{def.})}}c_{j'}e^{-itH_N^{(\text{def.})}} =\sum_{j}\left[e^{-it\tilde{h}}\right]_{j'j}c_{j}.
\end{align}
Together with Eq.~\eqref{eq:c_momentum}
we obtain
\begin{align}
\tilde{c}_{p}(t) & =\sum_{j}\left(\frac{1}{N}\sum_{j'}e^{-ij'p}\left[e^{-it\tilde{h}}\right]_{j'j}\right)c_{j} \\
 & =\sum_{j}\chi_{p,j}\left(t\right)c_{j}\label{eq:modes_mapping} \;.
\end{align}

For any sufficiently smooth deformation $g_j$, we can study the evolution in the continuum limit, $N\to\infty$ while keeping the length of the chain constant (see Appendix~\ref{app:continuum} for details). We show that, in the long-time limit, the PP evolution
maps all ``low-energy'' momentum modes $\tilde{c}_{p}(t)$, i.e, modes with
$p$ in the vicinity of the Fermi points $p=\pi/2$ and $p=3\pi/2$, to position space modes $c_{j}$ localized at the system's edges, i.e., with $j$ close to $j=1$ or $j=N$. In particular, for the parabolic deformation [Eq.~\eqref{parabolic_def}], the sum $\sum_{j=1}^{m}\left|\chi_{p,j}\right|^{2}$ 
for some $m<N$ and $p\approx\pi/2$ 
approaches unity exponentially [see Eq.~\eqref{eq:continuum_parab}], that is, the low-energy momentum modes quickly ``collapse'' to the edges. Once again, we note that a similar effect occurs for the SSD, albeit with an algebraic approach [see Eq.~\eqref{eq:continuum_SSD}].

The ``mode mapping'' provides further insight into the PP mechanism when one considers the dynamics of the correlator \mbox{$\left\langle \tilde{c}_{p}^{\dagger}\tilde{c}_{q}\right\rangle$} starting from the initial OBC GS $|\Omega_N\rangle$. The deviation of the correlator for the finite system of the size $N=36$ from the one for the infinite system,  \mbox{$\Delta\left\langle \tilde{c}_{p}^{\dagger}\tilde{c}_{q}\right\rangle =|\left\langle \tilde{c}_{p}^{\dagger}\tilde{c}_{q}\right\rangle -\left\langle \tilde{c}_{p}^{\dagger}\tilde{c}_{q}\right\rangle _{\infty}|$},
at times $t=0$ and $t=t_{\mathrm{PP}}$ is shown in Fig.~\ref{fig:correlators_mapping}(a) in momentum space and in Fig.~\ref{fig:correlators_mapping}(b) in position space, respectively.
We see that initially the deviation comes mainly from low-energy excitations with $p$ and $q$ being around the Fermi momenta $\pi/2$
and $3\pi/2$, see in Fig.~\ref{fig:correlators_mapping}(a).
Since the parabolically deformed Hamiltonian
does not perturb the true vacuum, the evolution only maps the initial deviation of the correlator in the momentum space to the deviation of the spatial correlators at the system boundaries, as shown in Fig.~\ref{fig:correlators_mapping}(b). 
This is illustrated in Fig.~\ref{fig:correlators_mapping}(c), where we show the product $X_p X_q$ with $X_p = \sum_{j \in A} \left|\chi_{p,j}\right|$, which quantifies the contributions of quasi-particles with initial momenta $q,~p$ to the two-point correlator.
The bulk is thus ``cleaned'' of excitations and provides a highly improved approximation of the true vacuum.

In practice, any initial state with relatively high energy density also exhibits
excitations with momenta far from the Fermi momentum. Since the equation of motion for these
quasi-particles is more complicated to solve analytically in the continuum limit
(see Appendix~\ref{app:continuum}), we studied numerically the momentum-position mapping function $\chi_{p,j}\left(t\right)$
[Eq.~\eqref{eq:modes_mapping}]. The left column of Fig.~\ref{fig:modes_mapping} shows $\left|\chi_{p,j}\left(t\right)\right|$ versus time $t$ and sites $j$ for fixed values of $p$ demonstrating the evolution of the position-space distribution for the selected modes $p$. We find that modes $p$ away from the Fermi momentum still collapse to a small region in position space, however, shifted towards the center of the system. We conclude that in the presence
of corresponding excitations at higher energies, the efficiency of the PP degrades in the sense that
the maximum size of an effectively cooled bulk decreases.

The right column of Fig.~\ref{fig:modes_mapping} shows the matrix $\left|\chi_{p,j}\left(t\right)\right|$
at fixed times where the corresponding modes $\tilde{c}_{p}$
shown in the left columns of the same figure are maximally localized in position space. These plots illustrate the spatial distribution of all momentum modes at the selected moments of time. We observe that
the optimal focusing time for the modes decreases for $p$ further away from the Fermi momentum. Thus, for a finite amount of initially populated momentum modes in the vicinity of the Fermi momentum, one can, in principle, optimize the time, where most of these modes are approximately localised in a finite edge region. This is demonstrated, i.e., by the middle plot of the right column of Fig.~\ref{fig:modes_mapping}.

As a summary, the mode mapping provides another interpretation of the PP dynamics, similar to the behaviour of light passing through an optical lens. While the above considerations have been explicitly discussed for a 1D chain, we present evidence in favor of the applicability of our PP protocol in two dimensions in Sec.~\ref{sec:2D} and Appendix~\ref{app:Hubbard}.

\subsection{Build-up of the Entanglement Hamiltonian, and verification of the target state
}
\label{section:EHapproach_for_tPP}

\begin{figure}
	\centering   
	\includegraphics[]{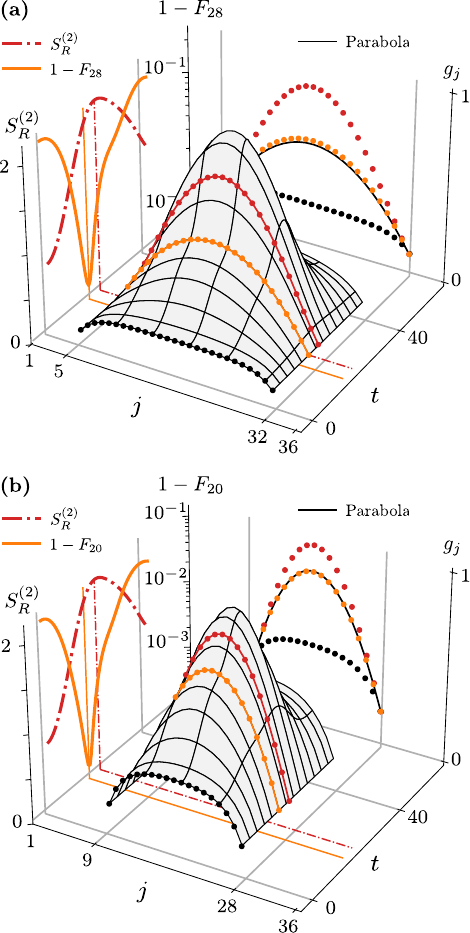}
	\caption{\label{fig:EH_build_up} Parameters $g^{(n)}_j$ of the learned EH $\tilde{H}_n = \sum_{j\in A} g^{(n)}_jh_{j,j+1}$ for subsystem $A$ consisting of (a) the sites $j = 5, \dots, 32$ ($n=28$) and (b) $j= 9, \dots, 28$ ($n=20$), starting from an initial thermal state (see text). The projections at the back show $g^{(n)}_j$ at times corresponding to the initial state (black dots with the lowest amplitude), maximum entropy (red dots with the maximum amplitude), and at the time of the infidelity minimum (orange dots) which are the closest to the parabolic shape (black solid line)}
\end{figure}

Since any reduced state $\rho_n \propto \exp (-\tilde{H}_n)$ is determined by the corresponding EH $\tilde{H}_n$, monitoring $\tilde{H}_n$ during the PP evolution allows us track the approach to the target state $\rho_n$. This becomes particularly feasible, e.g., at a critical point described by a CFT, where the Bisognano-Wichmann theorem predicts $\tilde{H}_n$ to be a parabolic deformation of the system Hamiltonian. To observe the build-up of this parabolic shape, we apply Hamiltonian learning~\cite{qi2019determining,Bairey2019} to find a local approximation $\tilde{H}_n=\sum_{j} g^{(n)}_j h_{j,j+1}$ of the EH~\cite{zhu2019reconstructing,kokail2021entanglement,kokail2021quantum} during the PP dynamics.

Fig.~\ref{fig:EH_build_up} shows the resulting EH parameters $g^{(n)}_j$, extracted from a numerical simulation of the PP protocol for two different subsystem sizes. We observe that the EH indeed becomes very close to the anticipated parabolic shape at an optimal time $t_\text{PP}$. This provides an experimentally accessible {\em verification} of the preparation of the desired target state with PP, and provides a way to identify the optimal time $t_\text{PP}$ in an experiment (see also below). For a discussion of Hamiltonian and EH learning protocols in an experimental context, we refer to~\cite{kokail2021entanglement,kokail2021quantum,carrasco2021theoretical}).

We emphasize that the results shown in Fig.~\ref{fig:EH_build_up} correspond to a {\em thermal} initial state $\propto e^{- H_N/T}$ at temperature $T=0.15$. This is in contrast to the {\em pure} initial states $\ket{\psi(0)}$ considered so far. Since we expect that any low-energy state evolves towards a purification of $\rho_n$, a mixture of low-energy states, such as a low-temperature thermal state, should also evolve to a mixture of purifications with the correct reduced density matrix. Our simulations support these expectations and demonstrate that the purity of the initial state is indeed not essential. Thus the present scheme achieves, with coherent quench dynamics, a `cooling of the bulk', i.e., a preparation of the reduced density matrix $\rho_n$ associated with the ground state of the infinite system.

We further note that the minimum of the infidelity (shown on the side of Fig.~\ref{fig:EH_build_up}) perfectly agrees with the time at which the reconstructed EH is closest to the parabolic shape. This is in contrast to the maximum of the R\'enyi entropy, which misses the optimal PP time due to the existence of excitations with higher energy in the initial thermal state. In the present case, where the structure of the expected EH is known, we can thus also estimate the optimal time $t_\text{PP}$ by minimizing the distance of the learned EH to the expected shape.

\subsection{\label{sec:2D}Fermions in 2D}
\begin{figure*}
\centering \includegraphics[]{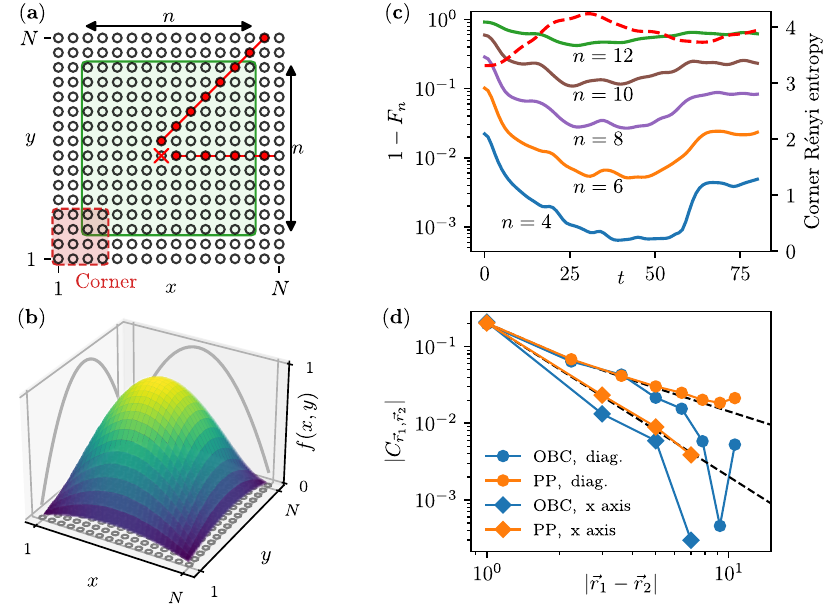} \caption{Purification Preparation for free fermions on a 2D square lattice.
\textbf{(a)} Schematic representation of the lattice with the linear size $N=16$ used in our numerical simulations.
The green solid rectangle indicates the bulk subsystem $n\times n$
with $n=12$. The red dashed rectangle shows the $4\times4$  corner region
chosen to measure the second R\'{e}nyi entropy $S^{(2)}$ during the PP. \textbf{(b)} Deformation $f(x,y)$ [Eq.~(\ref{eq:f_2d})]
used in the Hamiltonian [Eq.~(\ref{eq:H_PP_2d})] for the PP dynamics. Grey parabolic curves
shows projections of $f(x,y)$ onto the $x/y$ directions. 
\textbf{(c)} Evolution of the infidelity [Eq.~(\ref{eq:inF})] for
the bulk subsystems of several sizes $n$, and of the second R\'{e}nyi entropy $S^{(2)}$ in the corner region indicated in (a) during the PP dynamics. \textbf{(d)} Decay of two-point correlators [Eq.~(\ref{eq:cc_cors_r})] in the OBC and PP states along x axis and along a diagonal. $\vec r_1$ is fixed and is given by a red cross in (a). $\{\vec r_2\}$ are given in (a) by red bullets connected by red lines to indicate both correlators sets. Dashed lines give the correlators in the infinite state.
}
\label{fig:2D_fermi}
\end{figure*}

We conclude our investigation of PP for free fermions with an example in two spatial dimensions, considering a square lattice at half filling. The corresponding Hamiltonian reads
\begin{multline}
H_{N\times N}=\sum_{x=1}^{N-1}\sum_{y=1}^{N}\left(c_{x,y}^{\dagger}c_{x+1,y}+c_{x+1,y}^{\dagger}c_{x,y}\right)\\+\sum_{x=1}^{N}\sum_{y=1}^{N-1}\left(c_{x,y}^{\dagger}c_{x,y+1}+c_{x,y+1}^{\dagger}c_{x,y}\right) \;, \label{eq:2d_Fermi}
\end{multline}
where $c^{(\dagger)}_{x,y}$ denote fermionic annihilation(creation) operators at sites $(x,y)$ of a two-dimensional (2D) lattice as sketched in Fig.~\ref{fig:2D_fermi}(a).
Here, we consider a middle square of size $n\times n$ as subsystem $A$ (the `bulk')
as indicated by the green rectangle highlighted in Fig.~\ref{fig:2D_fermi}(a). 

For the PP dynamics, we choose the following deformation of the Hamiltonian given in Eq.~\eqref{eq:2d_Fermi},
\begin{multline}
H_{N\times N}^{(\mathrm{def.})}=\sum_{x=1}^{N-1}\sum_{y=1}^{N}f(x,y-\frac{1}{2})\left(c_{x,y}^{\dagger}c_{x+1,y}+c_{x+1,y}^{\dagger}c_{x,y}\right)\\+\sum_{x=1}^{N}\sum_{y=1}^{N-1}f(x-\frac{1}{2},y)\left(c_{x,y}^{\dagger}c_{x,y+1}+c_{x,y+1}^{\dagger}c_{x,y}\right),\label{eq:H_PP_2d}
\end{multline}
with $f(x,y)$ the product of two parabolas
along the $x$ and $y$ directions, i.e.,
\begin{equation}
f(x,y)=\frac{(N-x)x}{(N/2)^{2}}\times\frac{(N-y)y}{(N/2)^{2}},\label{eq:f_2d}
\end{equation}
see Fig.~\ref{fig:2D_fermi}(b). We have checked that an alternative deformation of parabolic type, namely
\mbox{$f(r,R=N/2)=\left(R-r\right)^{2}/R^{2}$} with $r$ the radius from
the subsystem's origin, yields similar results in the bulk. Once again, the parabolic deformation is suggested by the CFT version of the Bisognano-Wichmann theorem, which holds also in higher dimensions.
Here, we have chosen Eq.~\eqref{eq:f_2d} because it is better compatible with the square lattice (it is non-negative everywhere and zero at the boundary).

To detect the optimal time $t_{\text{PP}}$
to stop the PP, we track the second R\'{e}nyi entropy $S^{(2)}$~\citep{Brydges2019}, as in the 1D case considered before, but now at one corner of the
system shown in Fig.~\ref{fig:2D_fermi}(a) as the red dashed square. 
By analogy with the 1D case, we 
expect that a small corner (scaling sub-linearly with the full system size $N$) will suffice to detect $t_{\text{PP}}$.
For our simulations, we take $N=16$ and a $4\times4$ corner. 

Our results for the simulated PP protocol are presented in
Figure~\ref{fig:2D_fermi}(c), where we plot the evolution of the bulk infidelity starting from the ground states of the homogeneous finite-size
Hamiltonian [Eq.~\eqref{eq:2d_Fermi}]. Similar to the 1D case, we find that
the bulk infidelity for subsystems of different sizes $n$ with respect to the target reduced states of the 2D infinite system decreases significantly, demonstrating the success of PP. Moreover, the minima of the infidelity for different subsystem sizes approximately coincides with the first maximum of the second R\'{e}nyi entropy, measured in the small corner. At this point, the correlators of the subsystems approach the ones in the infinite system. Figure~\ref{fig:2D_fermi}(d) shows two-point correlators
\begin{equation}
     |C_{\vec r_1,\vec r_2}|\equiv|\langle c^\dagger_{\vec r_1}c_{\vec r_2}\rangle|,
     \label{eq:cc_cors_r}
 \end{equation}
with $\vec r_i=\{x_i,y_i\}$, that recover the target polynomial decay in the PP state in comparison with the initial ground state.

\section{\label{sec:int_spinchain}Extracting critical exponents}

\begin{figure}
\includegraphics[width=1\columnwidth]{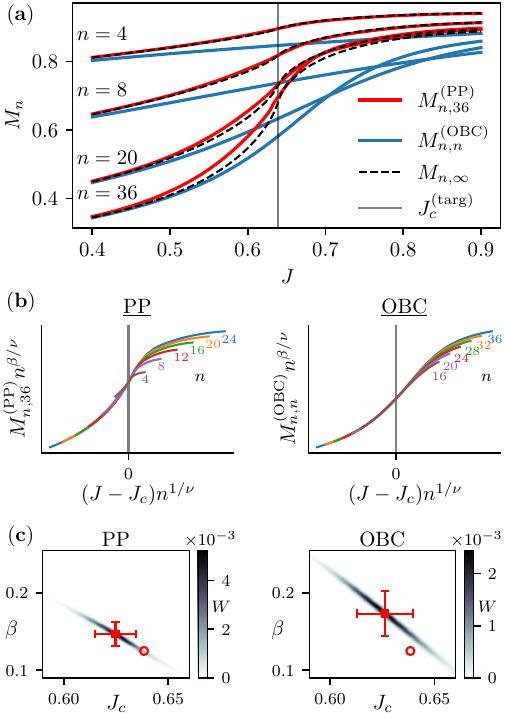} \caption{
\textbf{(a)} Root-mean-square staggered magnetization~(\ref{eq:magnetization})
for several subsystem sizes $n$ obtained from the OBC GS ($M_{n,n}^{(\text{OBC})}$),
from the PP states ($M_{n,N}^{(\text{PP})}$), and from the reduced
states of the infinite system ($M_{n,\infty}$), see the text. The grey solid
vertical line indicates the critical point $J_{c}^{(\text{targ})}$ numerically estimated by applying FSS analysis to iMPS states obtained by the iDMRG approach.
\textbf{(b)} Curves of the magnetization fitted to the scaling hypothesis~(\ref{eq:FSSA})
for the case $N=36$ and the average values $(J_{c},\beta)$ obtained
from the likelihood distribution in (c). The numbers give the subsystem
sizes $n$ for the curves with the corresponding colors. \textbf{(c)} Likelihood
distribution $W\left(J_{c},\beta\right)$. The red cross indicates the
average values of $(J_{c},\beta)$ with corresponding standard error. The purple
marker points the target values of $(J^{(\mathrm{targ})}_{c},\beta^{(\mathrm{targ})}=1/8)$.
}

\label{fig:magnetization}
\end{figure}

\begin{figure}
\includegraphics[width=1\columnwidth]{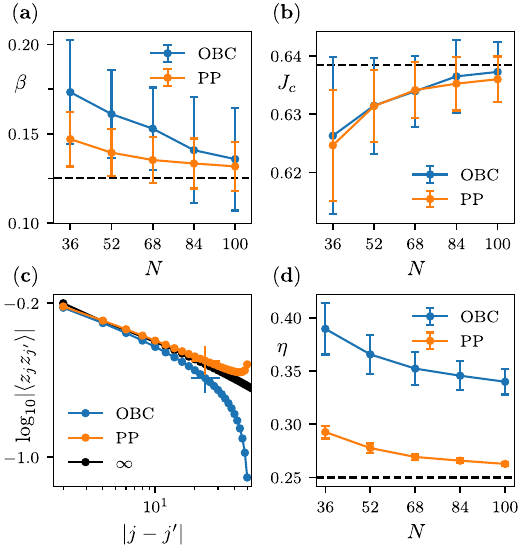} \caption{
\textbf{(a)} and \textbf{(b)} Convergence of the values $(J_{c},\beta)$ obtained from
the FSS analysis with growing sizes $N$ of the OBC GSs and the PP states.
The bars give the standard error. The dashed line indicates the target values. 
\textbf{(c)} The correlators $\left\langle z_{N/2}z_{j}\right\rangle $
in the OBC GS and the PP state of size $N=100$, and in the infinite
ground state, all at the average $J_{c}$ obtained from FSS with
the PP state of size $N=100$. The crosses mark the end of the range (at $j=N/4$) which was fitted with the polynomial decay Eq.~(\ref{eq:zz_fit}).
\textbf{(d)} Convergence of the retrieved critical exponent $\eta$ to the target value (dashed line) with
growing sizes $N$ of the OBC GSs and the PP states.
}
\label{fig:scaling}
\end{figure}

We now turn  to  application of PP to QPTs. Quantum critical phenomena can be grouped in universality
classes, which are characterized by a set of critical exponents and universal scaling functions~\citep{Blundell2016}.
Below we demonstrate that the usage of the PP bulk states for determining
critical exponents provides more accurate results than the direct usage of OBC GSs.

For this demonstration, we reconsider the interacting spin chain of Eq.\eqref{eq:SSH} and focus on the transition between the
trivial and AF phases along $\delta=3$ [see Fig.~\ref{fig:string_order}(a)]. Since the AF phase spontaneously breaks a $\mathbb{Z}_2$ symmetry ($\sigma^z \rightarrow - \sigma^z$), we expect critical behaviour of the 2D Ising universality class. In this case, the staggered magnetization serves as an order parameter, and for a finite system of size
$N$ we consider its root-mean-square (RMS)
\begin{equation}
M_{n}=\frac{1}{n}\sqrt{\left\langle \bigg(\sum_{j\in A}\left(-1\right)^{j}\sigma^z_{j},\bigg)^{2}\right\rangle }\label{eq:magnetization}
\end{equation}
in a subsystem $A$ of size $n\leq N$.

At criticality, we assume
the standard fine-size scaling (FSS) hypothesis
\begin{equation}
M_{n}\left(J\right)=n^{-\beta/\nu}\tilde{M}\left(n^{1/\nu}\left[J-J_{c}\right]\right),\label{eq:FSSA}
\end{equation}
with critical exponents $\beta$,$\nu$, scaling function $\tilde{M}$, and critical point $J_{c}$.
Note that in our scaling analysis, we use the size of the subsystem (bulk) $n$ rather than the total system size $N$.
For simplicity, in the following, we also use the exact value of $\nu=1$.

Figure~\ref{fig:magnetization}(a) shows the RMS of the order parameter $M_n$ for
several sizes $n$ accessible in a simulator with the relatively small
maximum number of spins $N=36$. For the PP states,
we evaluate $M_n$ in the reduced states $\text{Tr}_{\neg n}\left[\left|\psi(t_\text{PP})\right\rangle \left\langle \psi(t_\text{PP})\right|\right]$
of the PP state $\left|\psi(t_\text{PP})\right\rangle$ for $N$ sites, and the result is denoted as $M_{n,N}^{(\text{PP})}$. 
Employing the same strategy for the OBC GS $\left|\Omega_N\right\rangle$ (i.e., $M_{n,N}^{(\text{OBC})}$) suffers from strong boundary effects which invalidates the scaling hypothesis for the subsystem. 
Therefore, for a better comparison with the case of the OBC GS, we analyse $M_n$ in the OBC GSs $\left|\Omega_n\right\rangle$ of sizes $n$, i.e., $M_{n,n}^{(\text{OBC})}$. We see that the curves for $M_{n,N}^{(\text{PP})}$ are very close to those for the infinite system $M_{n,\infty}$, demonstrating also a steeper slope near the
phase transition as compared with $M_{n,n}^{(\text{OBC})}$, even for the maximum size $n=N$.

Using these data we obtain the location of the critical point $J_{c}$ and the critical exponent $\beta$ from a likelihood distribution $W\left(J_{c},\beta\right)$~\citep{Berges2014}
computed from the sets of the rescaled [according to Eq.~(\ref{eq:FSSA})] curves $M_{n,n}^{(\text{OBC})}$ and $M_{n,N}^{(\text{PP})}$ which are shown in Fig.~\ref{fig:magnetization}(b) (see Appendix~\ref{app:FSS} for details). Note that, in
order to reduce the finite-size effects for the OBC GS, we
exclude several smallest values of $n$, namely using only values $n=\{16,20,...,36\}$ accessible for the chosen $N=36$. For the PP
states, on the other hand, we exclude the ``hot'' boundaries by eliminating the same amount of the
largest values of $n$, namely using only $n=\{4,8,...,24\}$.
Figure~\ref{fig:magnetization}(c)
shows the corresponding likelihood distribution
$W\left(J_{c},\beta\right)$, the calculated average values of $J_{c}$ and $\beta$ with the corresponding error bars, and the target values. 
We see that the PP states result in a narrower likelihood distribution
$W\left(J_{c},\beta\right)$ and, therefore, in smaller uncertainty intervals for the average values, which are also closer to the target values (open circle).

We also investigate how the average values of $J_{c}$ and $\beta$, together with their uncertainties, change with increasing the maximum size $N$. The results are presented in Figs.~\ref{fig:scaling}(a) and (b), showing that the usage of the PP mainly improves the
accuracy and precision of the critical exponent $\beta$,
while the results for the critical value $J_{c}$ remains close to those obtained with the OBS GSs. 
However, we emphasize that in our analysis, we use the exact OBC GSs that are practically impossible to prepare with a quantum simulator.  Realistically, one  expects not the ground but a low-energy state
for which the accuracy of the FSS analysis might degrade due to the presence of the excitations. Applying our PP protocol to such imperfectly prepared GSs will push these excitations out of the bulk to the edges, making the FSS analysis insensitive to imperfections of the initial state preparation.

Finally, we consider the anomalous dimension $\eta$, which quantifies the deviation of the scaling dimension of the field from its engineering value. The quantity $\eta$ determines the polynomial decay of
the two-point correlator, which, for the considered critical point, has the form 
\begin{equation}
\left\langle \sigma^z_{m} \sigma^z_{j}\right\rangle \sim\left|m-j\right|^{-\eta}.\label{eq:zz_fit}
\end{equation}
We obtain the values of $\eta$ by fitting $\left\langle \sigma^z_{N/2}\sigma^z_{j}\right\rangle $
with \mbox{$j\in[N/2+1,3N/4]$} in the OBC GSs and in the PP states at the average
points $J_{c}$ retrieved from the FSS analysis with $M_{n,N}^{(\text{PP})}$.
The results for $N=100$, which are presented in Fig.~\ref{fig:scaling}(c), show that the correlator in the OBC GS decays faster than in the infinite
state at the same $J_{c}$, and its functional form deviates significantly from the algebraic one. In contrast, the correlators in the PP state approach the target values practically for all distances. As a result, $\eta$ obtained from the PP states with increasing $N$ converges much faster to the exact value, as demonstrated in Fig.~\ref{fig:scaling}(d). 

The above examples illustrate that the PP protocol improves the quantitative studies of quantum phase transitions on finite-size quantum simulators. It provides definite advantages in analyzing the scaling properties of the correlation functions and in extracting the critical indices.

\section{\label{sec:hubbard}Application: Fermi-Hubbard Models}
As an outlook for applications of our protocol to strongly correlated systems, we study the performance of PP for Fermi-Hubbard models (FHMs) for different lattice geometries and parameter regimes. In particular, we consider spinful fermions on a lattice, described by the Hamiltonian
\begin{align} 
\begin{split}
\label{eq:hubbard}
H_\text{FHM} = &-J \sum_{\braket{ij}, \sigma} \left( c_{i \sigma}^{\dagger} c_{j \sigma} + \text{H.c.} \right)\\ &+ U \sum_{i} n_{i \uparrow} n_{i \downarrow} - \mu \sum_{i \sigma} n_{i \sigma}. 
\end{split}
\end{align}
Here, the operators $c_{i \sigma}^{\dagger}$ ($c_{i \sigma}$) create (annihilate) a fermion on site $i$ with spin $\sigma \in \{ \uparrow, \downarrow \}$. The first line in Eq.~(\ref{eq:hubbard}) describes the hopping of particles between nearest-neighbour sites $\braket{ij}$ with transition amplitude $J$.  The second term causes a repulsive ($U>0$) onsite interaction if two fermions with opposite spin occupy the same site and the chemical potential $\mu$ controls the fermion filling in the ground state. Strongly correlated quantum phases of the FHM have been realized in quantum simulation experiments based on neutral atoms in optical lattices \cite{Mazurenko2017, hartke2020doublon, Vijayan2020}. Recent developments enable single-site addressing and allow for single-site readout of atoms (see \cite{NAP25613}). Moreover, spatial programmability of the Hamiltonian parameters, as required for PP, can be achieved by shaping optical potentials using digital mirror devices \cite{qiu2020precise}. As we will show below, PP enables to reveal the behavior of spatial correlation functions in the thermodynamic limit from experiments on very small lattice systems. 
\begin{figure}
\includegraphics[width=1\columnwidth]{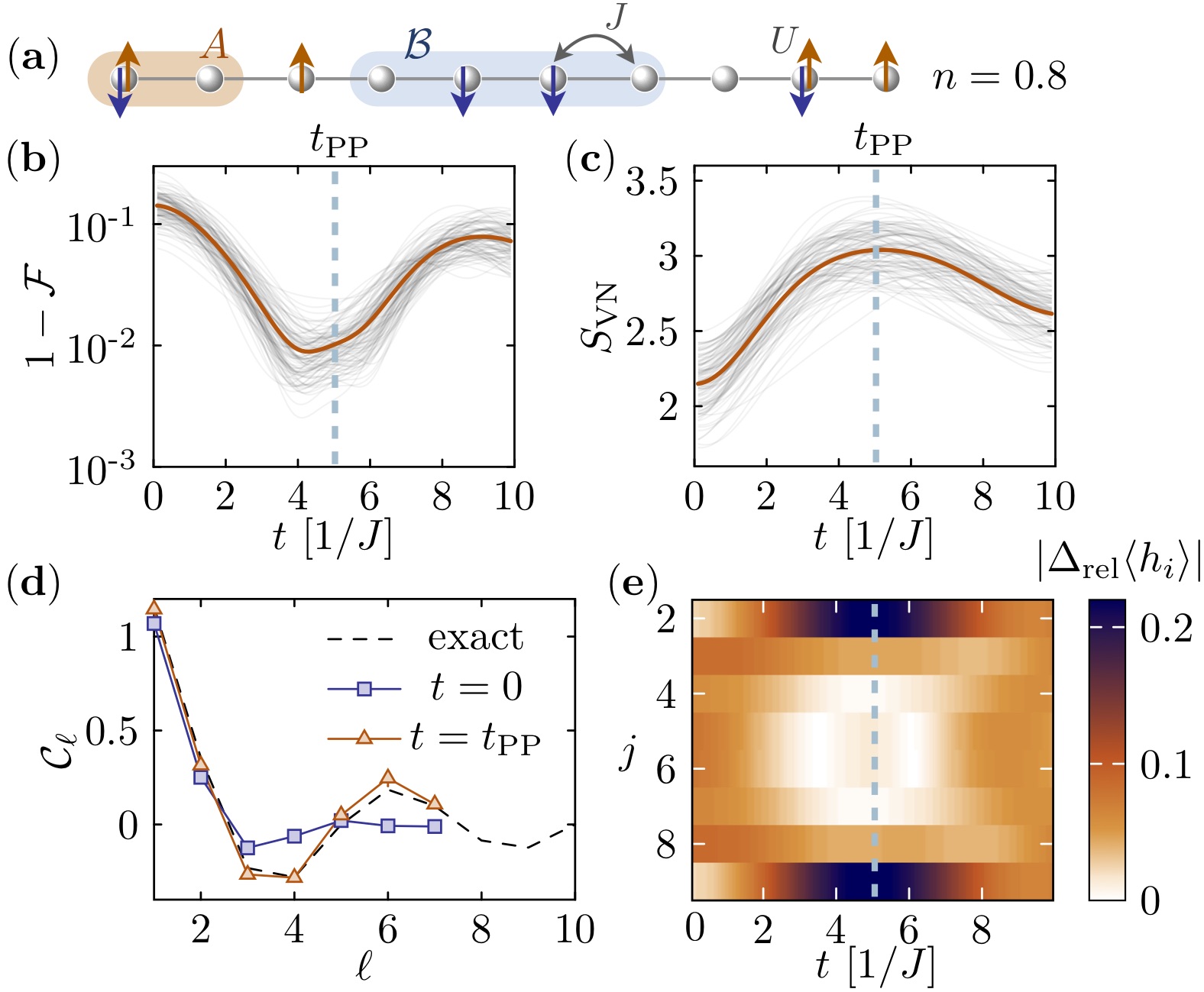} \caption{
Purification Preparation for a 10-site Hubbard model with $U/J=2$ and $40\%$ filling ($n = 0.8$). (\textbf{a}) Sketch of the Fermi-Hubbard chain with $n=0.8$.
(\textbf{b})
Fidelity [Eq.~(\ref{eq:inF})] computed between a bulk-subsystem ($\mathcal{B}$) of the 10-site chain and a corresponding subsystem in the ground state of a 200-site chain, during time-evolution with the deformed Hamiltonian. The grey lines show the results for 100 different { random superpositions} for the initial state (see main text). The bold solid line represents the mean over all experiments for different initial states.
(\textbf{c}) analogous plot to (\textbf{b}) for the Von Neumann entanglement entropy of the subsystem $A$ indicated in (\textbf{a}).
 (\textbf{d}) Decay of the 2-point correlation function $\mathcal{C}_{\ell} = 1/(N_\mathcal{B} - \ell) \sum_{i, \sigma} \left( \braket{c_{i\sigma}^{\dagger}c_{(i+\ell)\sigma}} + \text{c.c} \right)$ in the initial state and in the state at the optimal time $t_\text{PP}$. The black dashed line shows the decay of correlations in the ground state of a 200-site chain.
 (\textbf{e}) Absolute value of relative difference between local energy densities (see main text) for a 6-site bulk of the chain as a function of time.
}
\label{fig:hubbard}
\end{figure}

We start by studying the performance of PP for a 10-site Hubbard chain [see Fig.~\ref{fig:hubbard} (a)] with $U/J = 2$ and a small amount of hole doping $n = 0.8$. 
Here, we focus on a symmetry sector of zero magnetization with $\braket{N_{\uparrow}} = \braket{N_{\downarrow}} = 4$, where $N_{\sigma} = \sum_i n_{i \sigma}$ is the number of spin-$\sigma$ fermions. In order to demonstrate the robustness of our protocol, we take a {\em uniform random superposition} of the 10 lowest-lying eigenstates of $H_{\text{FHM}}$ as an initial state, which mimics an imprecise preparation of an initial state obtained e.g.~in non-adiabatic ground state preparation. The initial state is subsequently time evolved with spatially modified coupling parameters $J_i = J g_i$, $U_i = U g_{i-1/2}$ and $\mu_i = \mu g_{i-1/2}$, with $g_i$ given in Eq.~(\ref{parabolic_def}). This procedure is repeated 100 times to gather statistics for different initial state superpositions. 

Figure~\ref{fig:hubbard} summarizes the results obtained by averaging over the repetitions. Panel (b) shows the infidelity between a 4-site bulk-subsystem [Fig.~\ref{fig:hubbard}(a)] and a subsystem from the center of a 200-site chain. As can be seen, the infidelity decreases by more than one order of magnitude close to the optimal time $t_\text{PP}$, which approximately corresponds to 5 hopping times. Fig.~\ref{fig:hubbard}(c) demonstrated that the optimal time $t_\text{PP}$ is reflected in a maximum of the entanglement entropy for a 2-site subsystem at the boundary. In Fig.~\ref{fig:hubbard} panel (d) we plot the correlation function, analogously defined to Eq.~(\ref{eq:cc_cors}), for a $N_{\mathcal{B} = 8}$ site bulk subsystem of the 10-site chain at different times and for a bulk in the ground state of a 200-site chain (exact). Here, the 
correlators exhibit 
a rapid decay with distance modulated by oscillations with approximate 5-site periodicity. As can be seen in Fig.~\ref{fig:hubbard}(d), these modulations emerge during the quench with the deformed Hamiltonian and are accurately reproduced close to $t_\text{PP}$. Fig.~\ref{fig:hubbard}(e) shows the absolute relative error of local energy densities in the bulk-subsystem ($N_{\mathcal{B}} = 8$) of the 10-site chain as a function of time. The $h_i$ for a single site are given by
\begin{align}
\begin{split}
    h_i = &-\frac{J}{2} \sum_{\sigma} \left( c_{i-1 \sigma}^{\dagger} c_{i \sigma} + c_{i \sigma}^{\dagger} c_{i+1 \sigma} + \text{H.c.} \right)\\ &+ U n_{i \uparrow} n_{i \downarrow} - \mu \sum_{ \sigma} n_{i \sigma}, 
    \end{split}
\end{align}
with $c_{j \sigma}=0$ for $j\in\{0,N+1\}$, and the relative error 
\begin{equation}
\Delta_{\text{rel}}\left\langle h_i\right\rangle \equiv\left(\left\langle h_i\right\rangle -\left\langle h_i\right\rangle _{\text{200}}\right)/\left\langle h_i\right\rangle _{\text{200}},
\end{equation}
with $\left\langle h_i\right\rangle _{\text{200}}$ the energy density in the bulk of the 200-sites ground state. Note that $H_{\text{FHM}}$ of Eq.~(\ref{eq:hubbard}) is given by $H_{\text{FHM}} = \sum_i h_i$. Fig.~\ref{fig:hubbard}(e) demonstrates that close to the optimal time, $t_\text{PP}$, $\Delta_{\text{rel}}\left\langle h_i\right\rangle$ in the bulk vanishes, while a high energy density is accumulated at the edges.


As a first proof of principle demonstration for quasi-2D systems, we also studied the performance of PP on a $2\times5$ and $2\times6$-Hubbard ladder at $U/J = 2$, which is presented in Appendix~\ref{app:Hubbard}. As shown in Figs.~\ref{fig:hubbardladder} and ~\ref{fig:hubbardladder_2x6} of Appendix~\ref{app:Hubbard}, we observe a decreasing bulk infidelity with respect to large ladders and a corresponding improvement of local observables, providing preliminary evidence in favor of our protocol. 
We leave a more thorough study of PP on larger ladders, including an investigation of the sensitivity with respect to the Hamiltonian deformation, for future work.

\section{\label{sec:discussion}Discussion and Summary}

Quantum simulators represent engineered many-body systems isolated from the environment. Traditionally, analog quantum simulation addresses questions of realizing specific many-body Hamiltonians and preparing and studying corresponding equilibrium and non-equilibrium quantum phases  and dynamics. This includes the preparation of ground states or the study of quench dynamics in closed quantum systems \footnote {We assume that the quantum simulator naturally realizes a many-body system with {\em open} boundary conditions in spatial dimension $d$.}. In contrast, the present work outlines quantum protocols running on quantum simulators where the goal is to prepare {\em reduced density matrices}  as  {\em mixed}  states of many-body systems. The specific example considered in the present paper is the preparation of the reduced density matrix $\rho_n$ of a translation-invariant ground state of an {\em infinite} system on a quantum simulator,  where $n$ is the subsystem of interest and $N$ the size of the simulator ($n<N$). In this sense, a finite-size quantum simulator can `represent' the quantum state of an infinite system, in loose analogy to what is addressed with classical techniques such as iMPS or iPEPS~\citep{Schollwock2011, Orus2014}. 

The quantum protocols proposed and studied in the present paper are based on quantum memory of the simulator storing a {\em pure} state many-particle wavefunction $\ket{\psi}$, which represents a purification of the desired {\em mixed} target state $\rho_n$. The challenge addressed consists in devising coherent dynamics to prepare the target state of interest with given Hamiltonians available on the quantum simulator. As shown in this work, for ground states, this is achieved with coherent quench dynamics generated by {\em spatially deformed}  system Hamiltonians and starting from an approximate finite-size ground state, as realizable with state-of-the-art quantum simulators~\citep{NAP25613,altman2021quantum}. The specific choice of a parabolically deformed Hamiltonian is motivated by predictions of the Bisognano-Wichmann theorem on the structure of reduced density matrices for ground states, which then also provides a toolset to monitor the approach to the `correct' target state in an experiment.  The present paper has illustrated both robustness and broad applicability of this scheme of Purification Preparation in numerical simulations for a range of many-body models. 

The physical picture underlying the `cooling' to the target (ground) state $\rho_n$ in the `bulk' of the  simulator, with time evolution generated by a parabolically deformed system Hamiltonian, is best understood in a quasi-particle picture. Here, the coherent quench dynamics with the deformed Hamiltonian moves excitations to and accumulates them at the `edges' of the simulator, which play the role of a quantum reservoir. While, assuming preparation of an initial pure state, the overall dynamics remains in a pure state at all times, it is the entanglement (entropy) of bulk and edges which enables the preparation of the mixed target state $\rho_n$. As demonstrated with examples, this picture also generalizes to thermal quasi-particle excitations, i.e.~weakly mixed initial states. A promising  application is PP in Fermi-Hubbard models realized with cold atoms~\cite{Mazurenko2017,Koepsell2019}, which are currently limited by the experimentally available techniques to cool fermionic atoms.

Let us briefly comment on the general applicability of our method. Throughout this paper, we have found that a parabolic deformation of the system Hamiltonian, which is motivated by the BW theorem for a relativistic CFT with dynamical critical exponent $z=1$, is efficient in preparing the desired purification. However, based on the quasi-particle picture, we expect our protocol to be less efficient for transporting, e.g., low-momentum excitations with dispersion $\omega(p) \sim p^z$ with $z>1$. Similarly, it remains to evaluate the performance of our protocol in the presence of non-local excitations, e.g., in a topological system. In general, the validity of the quasi-particle picture in the presence of interactions could be tested by deriving a Boltzmann-type transport equation for general Hamiltonian deformations. We leave a thorough investigation of these issues for future work. In this context, it could be useful to include an additional step optimizing the shape of the deformation into our protocol.

We conclude by  commenting on the relation of Purification Preparation (PP)  to algorithmic cooling, and we refer to the review \cite{Park2016}, and experimental demonstrations in nuclear-magnetic-resonance based and ultracold atom platforms see \cite{Baugh2005} and \cite{Bakr2011}, respectively. The main idea of so-called heat-bath algorithmic cooling discussed in a quantum information context consists of extracting entropy from a subset of logical qubits and compressing it into ``reset'' qubits, which in turn release the entropy by coupling to a heat bath. While the compression step is superficially similar to the quasi-particle transport within our PP protocol, our goal is very different. In contrast to algorithmic cooling, which aims to prepare a fresh set of \emph{pure} qubits with minimal entropy, our target is a specific \emph{mixed} state corresponding to a part of the infinite-size ground state. It would be interesting to study the inclusion of dissipative elements into PP to iteratively extract the compressed quasi-particles from the boundaries of the system.

\section*{Acknowledgments}
We thank H. Pichler and R. van Bijnen for valuable discussions.  V.~K. was supported by the Austrian Research
Promotion Agency (FFG) via QFTE project AutomatiQ. This work was supported by European Union's Horizon 2020 research and innovation programme under Grant Agreement No.\ 817482 (Pasquans), and Simons Collaboration on Ultra-Quantum Matter, which is a grant from the Simons Foundation (651440, P.Z.). A.C.~acknowledges support from the Ministerio de Economía y Competividad MINECO (Contract No.   FIS2017-86530-P), from the European Union Regional Development Fund within the ERDF Operational Program of Catalunya (project QUASICAT/QuantumCat), from Generalitat de Catalunya (Contract No. SGR2017-1646), and from the UAB Talent Research program. The computational
results presented have been achieved (in part) using the HPC infrastructure
LEO of the University of Innsbruck.

\appendix
\section*{}

\section{\label{app:continuum}Continuum approximation in 1D}
The dynamics of the deformed free-fermion Hamiltonian in 1D of Sec.~III, 
\begin{align}
H = J \sum_n g_{n,n+1}\left( c_n^\dagger c_{n+1} + c_{n+1}^\dagger c_n\right) \;,
\end{align}
permits an illuminating continuum interpretation. To reveal this, we decompose the lattice operators as
\begin{align}
    c_n \sim e^{ik x} \psi_x = e^{i\bar{j} n} \psi_n \;, && \bar{j} = k a \;, && x=na\;,
\end{align}
where we assume that $\psi_x$ becomes a smooth functions in the limit $a\rightarrow 0$ for suitably chosen momentum $\bar{j}$. From the Heisenberg equation of motion, we have
\begin{align}
    i\partial_t \psi_n \sim J \left(e^{i\bar{j}}g_{n,n+1} \psi_{n+1} + e^{-i\bar{j}}g_{n-1,n} \psi_{n-1}\right) \;.
\end{align}
We now expand the field operators in derivatives up to second order,
\begin{align}
\psi_{n\pm 1} \sim \psi_x \pm a \partial_x \psi_x + \frac{a^2}{2} \partial_x^2 \psi_x \;,
\end{align}
and assume that an analogous expansion holds for the deformation
\begin{align}
    g_{n,n+1} \sim g_x + \frac{a}{2}\partial_x g_x + \frac{a^2}{8} \partial_x^2 g_x \;,\\
    g_{n-1,n} \sim g_x - \frac{a}{2}\partial_x g_x + \frac{a^2}{8} \partial_x^2 g_x \;.
\end{align}
Setting $J = v/(2a)$, we find
\begin{widetext}
\begin{align}
    i\partial_t \psi_x &\sim v \cos(\bar{j}) \left[\frac{g_x}{a} \psi_x +\frac{a}{2} \left(g_x \partial_x^2 \psi_x + (\partial_x g_x) (\partial_x \psi_x) + \frac{1}{4}\psi_x \partial_x^2 g_x\right)\right] + iv \sin(\bar{j}) \left[g_x \partial_x \psi_x + \frac{1}{2}\psi_x \partial_x g_x\right]\;,
\end{align}\label{eq:contff}
\end{widetext}
which approximately (to leading order in $a$) captures the operator evolution for quasi-particles around $\bar{j}$ under a smooth deformation $g_x$. Note that for $g_x=1$, we recover the usual left- and right-moving modes for a given fermion filling determined by $\bar{j}$, as it should be.

The above continuum approximation allows us to gain further insight for a given deformation $g_x$. In the following, we focus on the case of half filling and consider the right-moving modes at $\bar{j} = 3\pi/2$ (the left-movers only differ by some signs). Employing the method of characteristics to solve the partial differential equation for $\psi_x$, it follows that the quasi-particles travel on trajectories from $x_0$ to $x_1$ determined by
\begin{align}
    t(x_1,x_0) = \int_{x_0}^{x_1} \frac{dx}{v_x} \;, && v_x = v g_x 
\end{align}
with a position-dependent velocity $v_x$. 
We emphasize that the term $\propto \psi_x \partial_x g_x$ ensures a proper normalization of $\psi_x$, but it does not affect the trajectories. 
Indeed, for $\bar{j} = 3\pi/2$ $\eqref{eq:contff}$ is equivalent the Dirac equation for massless fermion propagating in a optical metric \cite{Boada2011}. Next, we compare the efficiency of the PP for a system of size $L$ evolved with a parabolic deformation or a SSD,
\begin{align}
g^{\text{par.}}_x = \frac{(L/2 + x)(L/2-x)}{(L/2)^2} \;, && g^{\text{SSD}}_x = \cos^2\left(\frac{\pi x}{L}\right) \;.
\end{align}
In these cases, the involved integrals can be evaluated analytically. For the parabolic deformation, we obtain the trajectories ($x_0 \neq -L/2$) as
\begin{align}\label{eq:parbolic_trajectory}
x^{\text{par.}}_1(t) &=  \frac{L}{2} \text{tanh}\left[\frac{2vt}{L} + \text{artanh} \left(\frac{2x_0}{L}\right)\right] \\
&\approx \frac{L}{2} \left(1- 2 \frac{1- (2x_0)/L}{1+ (2x_0)/L}e^{-4vt/L}\right) \;,
\end{align}
where the approximation is valid asymptotically for $t \gg L/(4v) \, \log \left[ (L-2x_0)/(L+2x_0)\right]$. Similarly, for the SSD, we obtain
\begin{align}
    x^{\text{SSD}}_1(t) &= \frac{L}{\pi} \text{arctan} \left[\frac{\pi v t}{L} + \text{tan} \left(\frac{\pi x_0}{L}\right)\right]\\
    &\approx \frac{L}{2} \left(1- \frac{2L}{\pi^2 v t}\right) \;,
\end{align}
where the asymptotic approximation applies for $t \gg  L/(\pi v) \, \text{tan} \left(\frac{\pi x_0}{L}\right)$ and $t\gg L/(\pi^2 v)$. This proves our claim made in the main text that the parabolic deformation is more efficient than the SSD in transporting the quasi-particles towards the edge, as the former shows an exponential approach $\propto \exp \left(-4vt /L\right)$, while the latter is only algebraical $\propto L/(v t)$.

These results also explain the mode mapping discussed in Sec.~\ref{sec:mode_mapping}, i.e., the transformation of initial lattice operators in momentum space to localized operators in position space at the edges of the system. Their reflection at times $t\gtrsim L/v$ is a lattice effect that happens when the $\psi_x$ operators approach a delta-like shape $\sim \delta(x\pm L/2)$ and the continuum approximation breaks down. Neglecting this reflection, we can estimate the weight $w_{\Delta x}(t)$ of the operator $c_{p=\bar{j}}(t)$ contained in an edge region of size $\Delta x$ by summing up all trajectories that have reached this region, i.e.,
\begin{align}
    w_{\Delta x}(t) = \frac{1}{L}\int_{-L/2}^{L/2} dx_0 \, \Theta\left(x_1(t)-L/2 +\Delta x\right) \;,
\end{align}
where $\Theta (\dots)$ is the Heaviside step function and we integrate of the initial conditions of the trajectory $x_1(t)$. For the above results for the parabolic and SSD, we see that the characteristic curves never cross, such that $w_{\Delta x}(t) = 1/2-x_0(t)/L$, where $x_0(t)$ is obtained by inverting $x_1(t) = L/2 -\Delta x$. Explicitly,
\begin{align}
    w^\text{par.}_{\Delta x}(t) &= \frac{1}{2} +\frac{1}{2} \text{tanh} \left[\frac{2vt}{L} + \text{atanh} \left(\frac{2\Delta x - L}{L}\right)\right]\label{eq:continuum_parab} \;,\\
    w^\text{SSD}_{\Delta x}(t) &= \frac{1}{2} +\frac{1}{\pi} \text{arctan} \left[\frac{\pi vt}{L} - \text{cot} \left(\frac{\Delta x \pi}{L}\right)\right]\label{eq:continuum_SSD} \;.
\end{align}
Given $w_{\Delta x}$, the probability density for finding a quasi-particle at position $x\in(-L/2,L/2)$ and time $t$ is given by $\left|\chi_x(t)\right|^2 = \left.\partial_{\Delta x} w_{\Delta x}\right|_{\Delta x = L/2+x}$. For comparison with the numerical results, we estimate the contribution of $c_{p=\bar{j}}(t)$ to a correlation function $\langle \cdots c_{p=\bar{j}}(t) \cdots \rangle$ in a subsystem $A = [x_a,x_b]$ as $X_{p=\bar{j}}(t) = \int_{x_a}^{x_b}dx \sqrt{\partial_x w_x}$. The explicit expressions are
\begin{align}\label{eq:continuum_parab2}
    X^{\text{par.}}_{p=3\pi/2}(t) &= \frac{\sqrt{L}}{2\sinh(2vt/L)}\log\left[\frac{L-2x_b + e^{4 v t /L}(L+2x_b)}{L-2x_a + e^{4 v t /L}(L+2x_a)}\right]\;.
\end{align}
Repeating the calculation for $p=\pi/2$ gives an analogous result for $X_{p=\pi/2}(t)$. For the initial OBC GS discussed in the main text, the dominant contribution comes from correlations between left- and right-movers. The error in corresponding two-point functions evaluated in the bulk $A$, originating from the initial correlator $\langle c^\dagger_{p=\pi/2} c_{q=3\pi/2}\rangle$, is thus estimated to decrease as the product $X_{p=\pi/2}(t) X_{p=3\pi/2}(t)$.
To compare these continuum predictions with the simulation results for the lattice model, we replace $L=Na$, $v=2Ja$ and $x_{a/b} = (n_{a/b}-N/2)a$, where $N$ is the number of lattice sites for a subsystem consisting of the sites \mbox{$\{n_a+1, n_a + 2, \dots , n_b-1, n_b \} \subset \{1,2,\dots,N-1,N\}$}. Specifically in Fig.~\ref{fig:correlators_mapping}(a)
we considered $N=36$ and $n_a=N-n_b = 4$. The analogous curve in Fig.~\ref{fig:correlators_mapping}(a) for the SSD involves integrals which we evaluated numerically.

\section{\label{methods:FF_simulation}Simulation with free fermions}

The simulations of free fermionic systems in the main text are performed semi-analytically by evolving the $N\times N$ two-points correlation matrices $\langle c_{j}^{\dagger}c_{m}\rangle$ that completely determine Gaussian states. This enables the calculation of GSs~\citep{Peschel2003,Wen2018}, thermal states~\citep{Magan2016}, the reduced infinite states~\citep{Peschel2009,Bahovadinov2019}, time evolution~\citep{Wen2018}, and computation of the Uhlmann fidelity between two Gaussian states~\citep{Amin2018}. To accurately compute fidelity and the R\'{e}nyi entropy from the $N\times N$ correlation matrices, we use up to $\sim2N$ decimal digits of precision.

\section{\label{app:FSS}Finite-size scaling analysis}

Here we provide more details on the finite-size scaling
analysis used to determine critical exponents in Sec.~IV of the main
text. For each maximum system size $N$, we compute the curves of
the staggered magnetization $M_{n}(J)$ versus the parameter $J$
in the region around a suspected critical point $J_{c}^{(\text{targ})}$
for several values of $n\le N$. The strategy to choose $n$ is given
in the main text. To obtain the critical exponent $\beta$ and the
critical point $J_{c}$ we fit $M_{n}(J)$ to the scaling hypotheses
\[
M_{n}\left(J\right)=n^{-\beta/\nu}\tilde{M}\left(n^{1/\nu}\left[J-J_{c}\right]\right),
\]
in which, for simplicity, we fix the exact value $\nu=1$. The fit
is implemented as follows:
\begin{enumerate}
\item Functions $\tilde{M}_{n}(n^{1/\nu}\left[J-J_{c}\right])=M_{n}(J)n^{\beta/\nu}$
computed at discrete $J\in[J_{\text{min}},J_{\text{max}}]$ are interpolated
on the interval. In our example we use $[J_{\text{min}},J_{\text{max}}]=[0.4,0.9]$,
with the target value $J_{c}^{(\mathrm{targ})}\approx0.6384$. 
\item The curve $\tilde{M}_{n_{\text{min}}}$ with the smallest $n=n_{\text{min}}$
is numerically parametrized by $s\in[0,1]$ such that the curve length is proportional $s$. For a given $J_{c}$ we obtain a corresponding
value $s_{c}$ and define an internal $[s_{\text{min}},s_{\text{max}}]$,
in our example we use $[s_{c}-0.1,s_{c}+0.1]$. 
\item Within the interval $[s_{\text{min}},s_{\text{max}}]$, we probe
$\tilde{M}_{n_{\text{min}}}(s)$ at $101$ homogeneously distributed
points $\{s_{i}\}$ such that $s_{0}=s_{\text{min}}$ and $s_{100}=s_{\text{max}}$.
At each point $\tilde{M}_{n_{\text{min}}}(s_{i})$ we find a normal
line (orthogonal to the tangent) to the curve at this point, and then numerically obtain points
$\{s_{i}^{\left(n\right)}\}$ at which the normal line crosses all
other curves $\tilde{M}_{n}$.
\item Using this data we define a cost function which describes the collapse
of the rescaled curves along the probe interval $[s_{\text{min}},s_{\text{max}}]$
as
\[
\varepsilon^{2}(J_{c},\beta)=\frac{\sum_{s_{i},n}\left[\text{Std}(\{\tilde{M}_{n}(s_{i}^{(n)})\})^{2}+\text{Std}(\{s_{i}^{(n)}\})^{2}\right]}{\sum_{s_{i}}\left[(\overline{\{\tilde{M}_{n}(s_{i}^{(n_{\text{min}})})\}})^{2}+(\overline{\{s_{i}^{(n_{\text{min}})}\}})^{2}\right]},
\]
with $\text{Std}$ the standard error and the horizontal bar stands for
the arithmetic mean. 
\item The cost function $\varepsilon^{2}(J_{c},\beta)$ is used to define
the likelihood distribution~\citep{Berges2014}
\[
W(J_{c},\beta)=\text{exp}\left[-\frac{\varepsilon^{2}(J_{c},\beta)}{2\varepsilon_{\text{min}}^{2}}\right]
\]
with $\varepsilon_{\text{min}}^{2}$ the smallest value among the
probed points $(J_{c},\beta)$. From $W(J_{c},\beta)$ we obtain the
average values as
\begin{align*}
\left\langle J_{c}\right\rangle  & =\sum_{J_{c},\beta}J_{c}W(J_{c},\beta),\\
\left\langle \beta\right\rangle  & =\sum_{J_{c},\beta}\beta W(J_{c},\beta),
\end{align*}
and the standard errors
\begin{align*}
\Delta\left\langle J_{c}\right\rangle  & =\sqrt{\sum_{J_{c},\beta}\left(J_{c}-\left\langle J_{c}\right\rangle \right)^{2}W(J_{c},\beta)},\\
\Delta\left\langle \beta\right\rangle  & =\sqrt{\sum_{J_{c},\beta}\left(\beta-\left\langle \beta\right\rangle \right)^{2}W(J_{c},\beta)}.
\end{align*}
\end{enumerate}

\section{Trotterized Purification Preparation}\label{app:Trotter}

\begin{figure}
\includegraphics{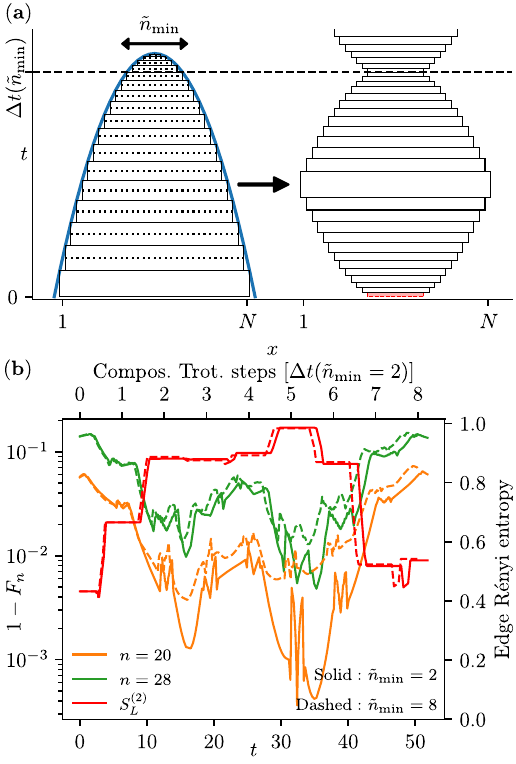}

\caption{
\textbf{(a)} Scheme to design the Trotterized evolution from the deformation
given by the blue solid curve, see the text. 
\textbf{(b)} Trotterized
PP of the $N=36$ 1D system of free fermions starting from the OBC
GS. Trotterizatios with $\tilde{n}_{\text{min}}=2$ (solid lines)
and $\tilde{n}_{\text{min}}=8$ (dashed lines) are given. The orange
and green lines show the infidelity and the red lines give the R\'{e}nyi entropy $S^{(2)}_L$
obtained on the two most left sites.
}
\label{fig:Trot_evol}
\end{figure}

In some experimental systems, such as neutral atoms in optical lattices or 
Rydberg atoms in tweezer arrays the deformed Hamiltonian required for the PP 
can be directly implemented by tuning the couplings locally.
Here we outline an alternative implementation of the PP dynamics based on
the Trotterization of the time evolution of the deformed Hamiltonian.
Such a Trotterized evolution,
\[
U_{\text{trot}}=\prod_{\text{Compos}.\,\text{Trot}.\,\text{steps}}\prod_{\tilde{n}}e^{-i(\delta t)_{\tilde{n}}H_{\tilde{n}}} \;,
\]
uses {\em homogeneous} Hamiltonians $H_{\tilde{n}}$ applied to successively
growing or decreasing subsystems of size $\tilde{n}$, as sketched
in Fig.~\ref{fig:Trot_evol}(a) where we consider Trotterization
of the parabolic deformation [Eq.~\eqref{parabolic_def}]. In the plot,
each box corresponds to an elementary Trotter steps with width $\tilde{n}$
and height $2(\delta t)_{\tilde{n}}$. Because of the experimental
limitations, we assume the shortest time step accessible in the experiment
from which one can obtain the smallest size $\tilde{n}_{\text{min}}$.
To achieve a Trotter error that scales quadratically with the Trotter steps, 
we halve the rectangles along
the time axis and reorder them as shown in the figure. 
We also move the first elementary step (red rectangle), having the shortest evolution
time, to the end of the sequence and stacked it with the identical
last step. This only affects the very first elementary step. The whole procedure
gives the Trotter-Suzuki expansion of the PP at second order in the composite step $\Delta t(\tilde{n}_{\text{min}})$.

In Fig.~\ref{fig:Trot_evol}(b), we give the result of the Trotterized
PP for a $N=36$ system of free fermions starting from the OBC GS.
We consider $\tilde{n}_{\text{min}}=2$ and $\tilde{n}_{\text{min}}=8$
cases, where we fix the time step to $(\delta t)_{2}=0.01$, where we measure 
time in units of the inverse coupling strength in the original Hamiltonian. 
As a result of the Trotterization scheme, for $\tilde{n}_{\text{min}}=2$, the smallest
Trotter step becomes $2(\delta t)_{2}=0.02$ (stacked elementary step
between two composite steps), and, for $\tilde{n}_{\text{min}}=8$,
the smallest elementary step takes time $(\delta t)_{10}=0.09$. In
both cases, the minima of the infidelity determines optimal times close
to the one obtained with the analog evolution, Fig.~\ref{fig:PP_of_1D_fermi}(c).
In the considered scenarios, the optimal evolution takes $5$ composite
Trotter steps equivalent to $170$ and $140$ elementary trotter steps
in the case of $\tilde{n}_{\text{min}}=2$ and $\tilde{n}_{\text{min}}=8$, 
respectively. Notice that the latter can be obtained from the former by eliminating the smallest middle steps. By doing so, we effectively realize Trotterization
of a different deformed Hamiltonian with a plateau in the bulk, which explains the degradation
in the efficiency of the PP observed for $\tilde{n}_{\text{min}}=8$.

\section{Purification Preparation for Fermi-Hubbard Models: Additional Results}\label{app:Hubbard}

Here we provide additional results for applying our protocol for Purification Preparation (PP) to Fermi-Hubbard systems in quasi-2D and 1D geometries. In particular, we discuss the performance of PP for small Hubbard ladders with hole doping, which provides a first proof of principle demonstration of our protocol for a quasi-2D system in a strongly correlated regime. Furthermore, we present additional numerical results for Hubbard chains for different coupling parameters compared to the one presented in the main text. 

\subsection{Fermi-Hubbard Model in a Two-Leg Ladder Geometry}

\begin{figure}
\includegraphics[width=1.0\columnwidth]{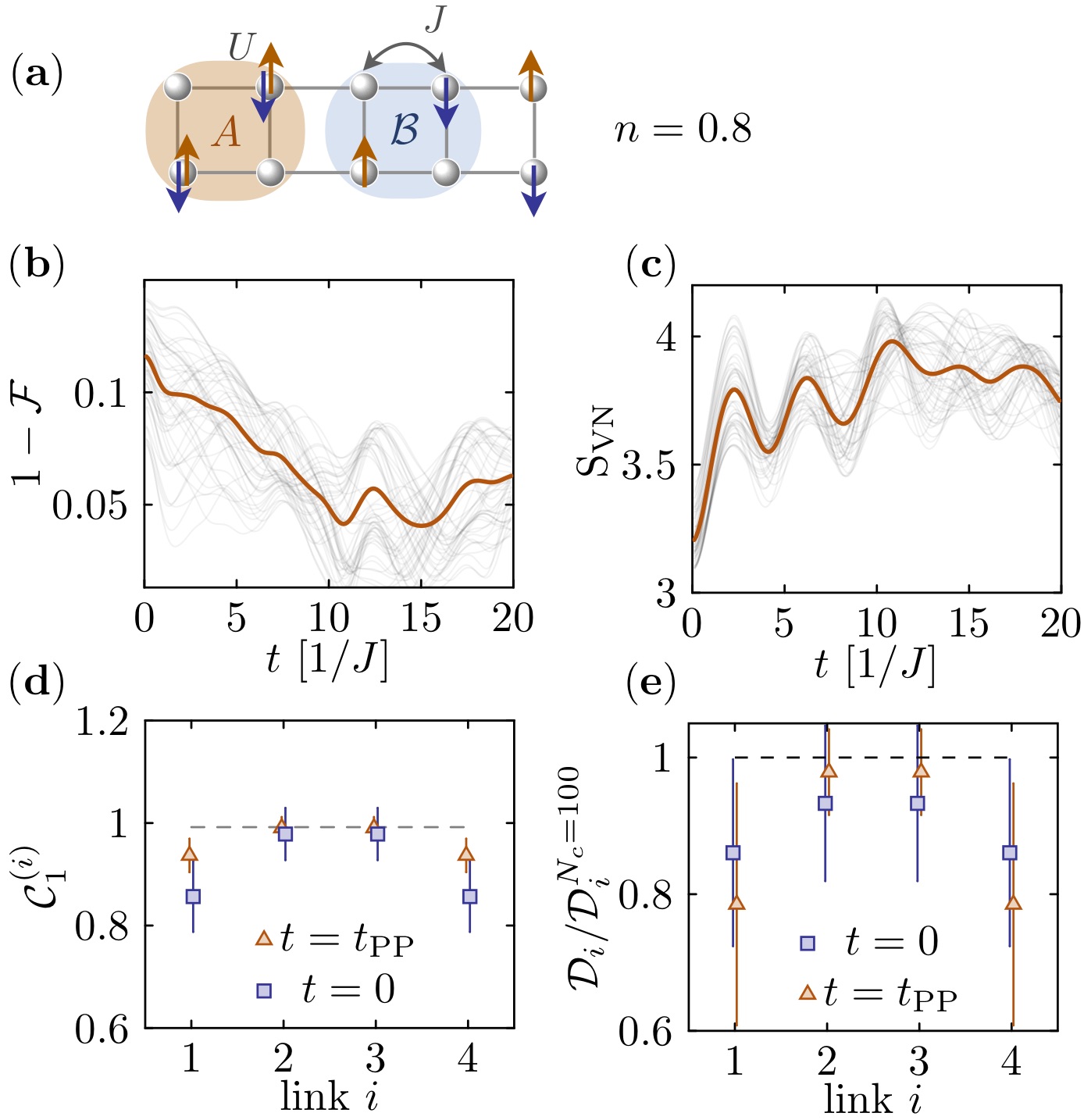} \caption{Results for PP on a Hubbard ladder (\textbf{a}) Sketch of the $2\times5$ Fermi-Hubbard ladder at $n=0.8$. (\textbf{b}) Fidelity [Eq.~(10) in main text] computed between a bulk-subsystem ($\mathcal{B}$) [see (\textbf{a})] and a corresponding subsystem in the ground state of a $2\times100$ ladder, during time-evolution with the deformed Hamiltonian. The grey lines show the results for 50 different initial states given by a uniform random superposition of the two lowest Hamiltonian eigenstates. The bold solid line represents the mean over all experiments for different initial states. (\textbf{c}) Analogous plot to (\textbf{b}) for the Von Neumann entanglement entropy of the subsystem $A$ indicated in (\textbf{a}).
(\textbf{d}) Average nearest-neighbor hopping correlation function in the $x$-direction $\mathcal{C}_1^{(i)} = \sum_{\sigma} \braket{c_{i, 1, \sigma}^{\dagger} c_{(i + 1), 1, \sigma} } + \text{c.c.} $, computed in the upper leg of the ladder at the initial ($t=0$) and the optimal time ($t = t_\text{PP}$). The dashed line represents the value from the center of a $2\times100$ ladder. (\textbf{e}) Nearest-neighbor $d$-wave pair correlations $\mathcal{D}_i = \text{Re} \braket{\Delta_i^{\dagger} \Delta_{i+1}}$ at $t=0$ and $t=t_\text{PP}$. The plot shows the ratio between $\mathcal{D}_i$ and $\mathcal{D}_i^{N_c=100}$ obtained from averaging over a small bulk-region in the central region of the ground state of a $2\times100$ ladder.
}
\label{fig:hubbardladder}
\end{figure}

We investigate the performance of PP for a $2\times5$-Hubbard ladder [see Fig.~\ref{fig:hubbardladder} (a)] described by the Hamiltonian Eq.~(22) in the main text. We study the system at $40\%$ filling, i.e. $\braket{N_\uparrow} = \braket{N_\downarrow} = 4$ and at $U/J = 2$.

In our numerical experiments we initialize the system in a random superposition of the two lowest-lying energy eigenstates. This initial state is time evolved with a suitable deformation of the Hamiltonian [Eq.~(22) in main text] adopted to ladder systems:

\begin{align}\label{eq:hubbarddef}
\begin{split}
    H_{\text{FHM}}^{\text{(def.)}} = &-J \sum_{i, \lambda, \sigma} g_i \left( c_{i,\lambda, \sigma}^{\dagger} c_{(i + 1),\lambda, \sigma} + \text{H.c.} \right) \\
    &-J \sum_{i, \sigma} g_{i - \frac{1}{2}} \left( c_{i,1, \sigma}^{\dagger} c_{i,2, \sigma} + \text{H.c.} \right) \\
    &+U \sum_{i, \lambda} g_{i - \frac{1}{2}} n_{i,\lambda, \uparrow} n_{i,\lambda, \downarrow} \\
    &+ \mu \sum_{i, \lambda, \sigma} g_{i - \frac{1}{2}} n_{i, \lambda, \sigma}
    \end{split}
\end{align}
with 
\begin{align}
    g_i = \frac{i (N_c -i)}{(N_c/2)^2}
\end{align}
and $N_c$ the number of ladder rungs. The index $\lambda \in \{1,2\}$ denotes the two legs of the ladder. Note that in the definition of the deformed Hamiltonian Eq.~(\ref{eq:hubbarddef}) the index $i$ in the first line runs from $0$ to $N_c-1$, while it runs up to $N_c$ in the remaining lines. The coefficients $g_{i-1/2}$ are thus required to generate a symmetric parabola in the $x$-direction. We repeat the quench simulations 50 times in order to collect statistics for different random initial states. We emphasize that by sampling over random initial superpositions we demonstrate PP for two types of typical experimental imperfections. Since a single realization corresponds to a coherent superposition with of low-lying excited states, it models an imperfect preparation of the ground state, e.g. via almost adiabatic state preparation. The collection of all such realizations realizes a mixed state, akin to a microcanonical ensemble with a small energy window above the ground state. This ensemble thus models another type of experimental errors, similar to a finite (low) temperature. In either case, our simulations indicate the PP works well for these states ``close'' to the ground state. 

In Fig.~\ref{fig:hubbardladder}(b) we plot the Uhlmann-infidelity between the subsystem $\mathcal{B}$ depicted in Fig.~\ref{fig:hubbardladder}(a) and a corresponding bulk-subsystem in the ground state of a $2\times100$-ladder. We observe a continuous decrease of the infidelity, reaching a minimum at about $\sim$10 hopping times. Fig.~\ref{fig:hubbardladder}(c) demonstrates that the time of optimal infidelity is approximately reflected by a maximum in the entanglement entropy for a boundary subsystem $A$ shown in Fig.~\ref{fig:hubbardladder}(a).   

We now turn to analyzing local observables at different times during the evolution with the deformed Hamiltonian. Fig.~\ref{fig:hubbardladder}(d) shows nearest-neighbor hopping-correlations averaged over initial states ($t=0$) and over states at the optimal time ($t=t_\text{PP}$). As can be seen, correlations close to the system boundary are closer to the exact value at $t=t_{\text{PP}}$. While the bulk-correlators are already close to the exact value at $t=0$, the correlators at $t=t_\text{PP}$ have a significantly reduced error bar, implying that the exact correlations are consistently reproduced close to the optimal time. 

Finally, we investigate the behaviour of nearest-neighbor $d$-wave pair correlations during the time evolution with the deformed Hamiltonian (\ref{eq:hubbarddef}). For strongly correlated systems of Fermions, the decay of the $d$-wave pair correlation function can serve as a detector for pairing phases \cite{dolfi2015pair}. Thus, it plays an important role for characterising low-temperature phases in Fermi-Hubbard models. For a 2-leg ladder,  the $d$-wave pair correlation function is defined as 
\begin{align}
    \mathcal{D}_{i,j} = \braket{\Delta_i^{\dagger} \Delta_j }
\end{align}
where
$\Delta_i^{\dagger} = c_{i, 1, \uparrow}^{\dagger} c_{i, 2, \downarrow}^{\dagger} - c_{i, 1, \downarrow}^{\dagger} c_{i, 2, \uparrow}^{\dagger}$
creates a singlet on rung $i$ \cite{dolfi2015pair}. In Fig.~\ref{fig:hubbardladder}(e) we plot the real part of the averaged nearest-neighbor $d$-wave pair correlations $\mathcal{D}_i = \text{Re}\braket{\Delta_i^{\dagger} \Delta_{i+1} }$ for all $4$ links of the $2\times5$ ladder. For comparison, the data are divided by the averaged value of $\mathcal{D}_i$ from a 10-site bulk region of a $2\times100$ ladder.  Similar to the hopping correlators discussed above, we obtain large fluctuations for the correlators close to the boundary, while the mean value of the correlators in the center approaches the exact value with a significantly reduced error bar.

We amphasize that the results presented in this chapter where carried out for a small ladder system ($2 \times 5$) and thus may be severely affected by finite-size effects. Future investigations will have to include a more sophisticated analysis for larger ladders in several different parameter regimes as well as an investigation on how strongly the results depend on the type of Hamiltonian deformation.

\subsection{PP on a 2$\times$6 Fermi-Hubbard ladder}
\begin{figure}
\includegraphics[width=1.0\columnwidth]{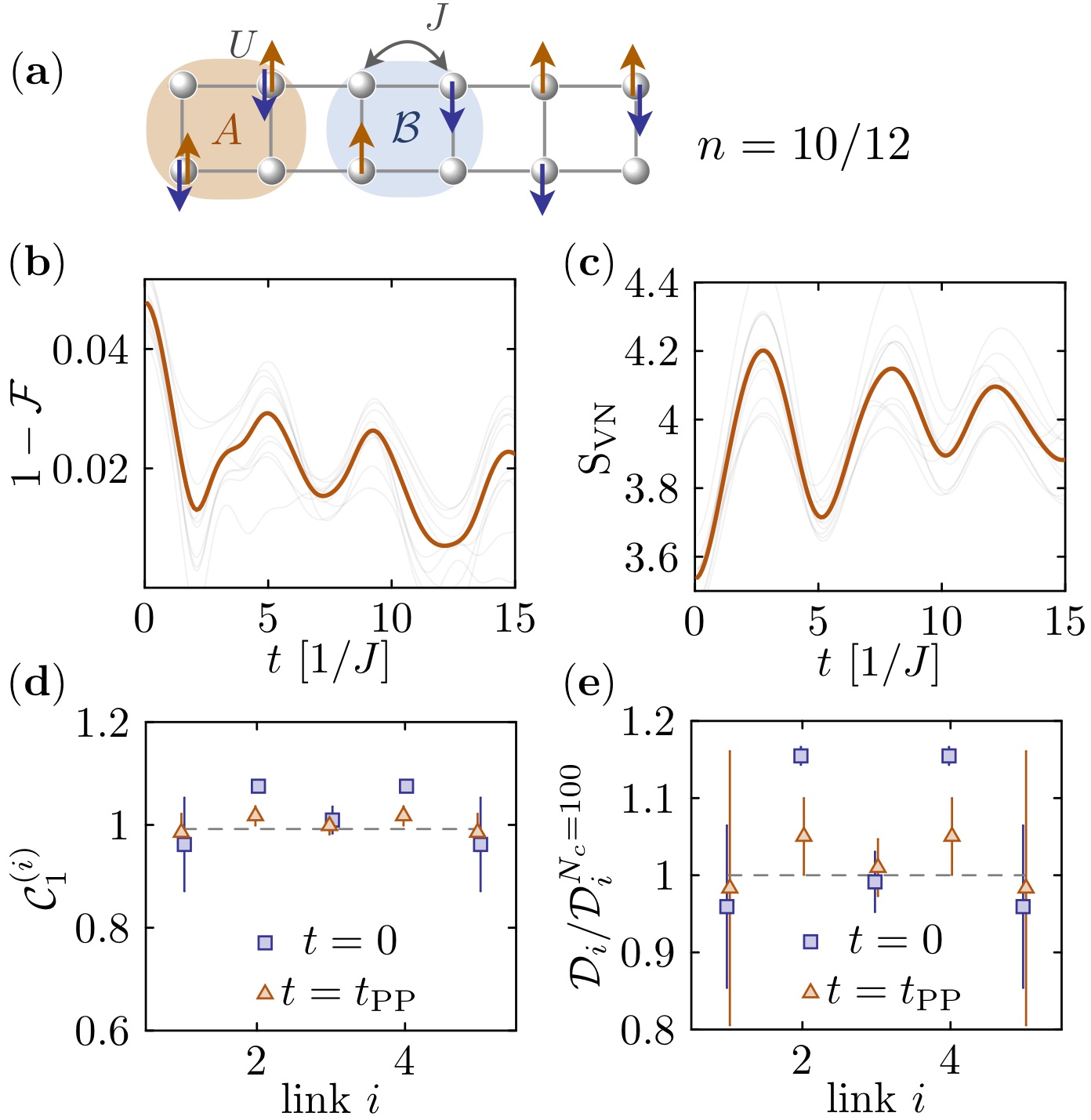} \caption{ Analogous plots to Fig.~\ref{fig:hubbardladder} for a $2\times$6 Fermi-Hubbard ladder. The initial state is a random superposition of the 2 lowest-lying energy eigenstates. The quench is repeated 10 times to gather statistics for different random initial states.
}
\label{fig:hubbardladder_2x6}
\end{figure}
In this section we repeat the simulations presented in Fig.~\ref{fig:hubbardladder} for a $2\times6$ strongly repulsive Fermi-Hubbard ladder. While the filling is chosen to be $n=10/12$ [see Fig.~\ref{fig:hubbardladder_2x6}(a)], the quench parameters are exactly the same as described in the previous paragraph. Again, we observe a significant drop of infidelity and corresponding nearest-neighbor correlation functions approaching the values in the $N_c=100$ ladder.

\subsection{Fermi-Hubbard Chain at Half-Filling}

\begin{figure}
\includegraphics[width=1\columnwidth]{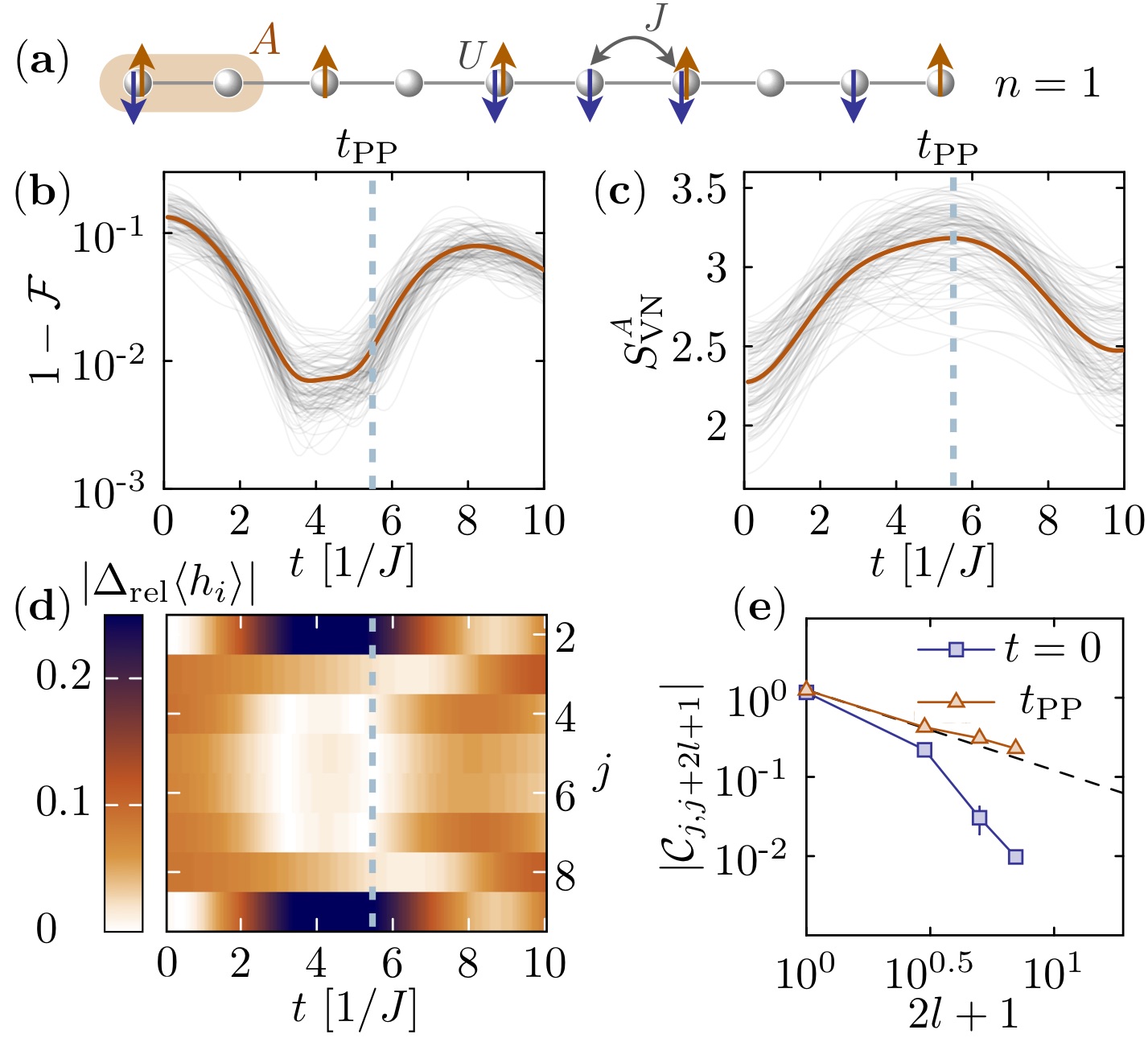} \caption{
Purification Preperation for a 10-site Hubbard model with $U/J=1$ and $50\%$ filling ($n = 1$). (\textbf{a}) Sketch of the Fermi-Hubbard chain with $n=1$.
(\textbf{b})
Fidelity [Eq.~(10) in main text] computed between a bulk-subsystem ($\mathcal{B}$) of the 10-site chain and a corresponding subsystem in the ground state of a 200-site chain, during time-evolution with the deformed Hamiltonian. The grey lines show the results for 100 different random superpositions for the initial state (see main text). The bold solid line represents the mean over all experiments for different initial states.
(\textbf{c}) analogous plot to (\textbf{b}) for the Von Neumann entanglement entropy of the subsystem $A$ indicated in (\textbf{a}).
 (\textbf{d}) Local energy densities $\braket{\hat{h}_i}$ (see main text) for a 8-site bulk of the chain as a function of time.
  (\textbf{e}) Decay of the 2-point correlation function $\mathcal{C}_{j, j+2l + 1} = \sum_\sigma \left( \braket{c_{j\sigma}^{\dagger}c_{(j+2l+1)\sigma}} + \text{c.c} \right)$ in the initial state and in the state at the optimal time $t_\text{PP}$. The black dashed line shows the decay of correlations in the ground state of a 200-site chain.
}
\label{fig:hubbardhalf}
\end{figure}

In this section we provide additional results for 1D Hubbard chain in a different parameter regime. The setting is equivalent to the one discussed in section V in the main text with the difference that we study the system at half filling $\braket{N_\uparrow} = \braket{N_\downarrow} = 5$ and $U/J = 1$. Overall, the performance of PP is very similar to the situation discussed in Fig.~10 in the main text. Fig.~\ref{fig:hubbardhalf}(e) shows that the power-law for the decay of correlations is accurately reproduced close to the optimal time $t=t_\text{PP}$ which we define to be at the maximum of the entanglement entropy Fig.~\ref{fig:hubbardhalf}(c).

\bibliographystyle{unsrt}
\bibliography{bibliography}

\end{document}